\documentclass[journal,onecolumn,12pt]{IEEEtran}
\usepackage{cite}

\ifCLASSINFOpdf
\else
\fi
\usepackage[cmex10]{amsmath}

\usepackage{psfrag,graphicx}

\begin{document}
%

\title{On the Gain of Joint Processing of\\ Pilot and Data Symbols in Stationary\\ Rayleigh Fading Channels}

\author{Meik~D\"orpinghaus,~\IEEEmembership{Member,~IEEE,}~Adrian~Ispas,\\and~Heinrich~Meyr,~\IEEEmembership{Life Fellow,~IEEE} 
\thanks{This work has been supported by the UMIC (Ultra High Speed Mobile Information and Communication) research centre. The material in this paper was presented in part at the 2010 International Zurich Seminar on Communications, Zurich, Switzerland, March 2010.}
\thanks{M. D\"orpinghaus was with the Institute for Integrated Signal Processing Systems, RWTH Aachen University, 52056 Aachen, Germany and is now with the Institute for Theoretical Information Technology, RWTH Aachen University, 52056 Aachen, Germany (e-mail: doerpinghaus@ti.rwth-aachen.de).}
\thanks{A. Ispas and H. Meyr are with the Institute for Integrated Signal Processing Systems, RWTH Aachen University, 52056 Aachen, Germany (e-mail: \{ispas,meyr\}@iss.rwth-aachen.de).}}

\maketitle


\maketitle

\begin{abstract}
In many typical mobile communication receivers the channel is estimated based on pilot symbols to allow for a coherent detection and decoding in a separate processing step. Currently much work is spent on receivers which break up this separation, e.g., by enhancing channel estimation based on reliability information on the data symbols. In the present work, we evaluate the possible gain of a joint processing of data and pilot symbols in comparison to the case of a separate processing in the context of stationary Rayleigh flat-fading channels. Therefore, we discuss the nature of the possible gain of a joint processing of pilot and data symbols. We show that the additional information that can be gained by a joint processing is captured in the temporal correlation of the channel estimation error of the solely pilot based channel estimation, which is not retrieved by the channel decoder in case of separate processing. In addition, we derive a new lower bound on the achievable rate for joint processing of pilot and data symbols.\looseness-1
\end{abstract}


\IEEEpeerreviewmaketitle

\section{Introduction}
\IEEEPARstart{V}{irtually} all practical mobile communication systems face the problem that communication takes place over a time varying fading channel whose realization is unknown to the receiver. However, for coherent detection and decoding an estimate of the channel fading process is required. For the purpose of channel estimation usually pilot symbols, i.e., symbols which are known to the receiver, are introduced into the transmit sequence. In conventional receiver design the channel is estimated based on these pilot symbols. Based on these channel estimates, in a separate step coherent detection and decoding is performed. Both processing steps are executed separately. 

In recent years, much effort has been spent on the study of iterative joint channel estimation and decoding schemes, i.e., schemes, in which the channel estimation is iteratively enhanced based on reliability information on the data symbols delivered by the decoder, see, e.g., \cite{Valenti2001,NoelsLottici2005,Godtmann2007,SchmittMeyrZhang09TCOMAC}. In this context, the channel estimation is not solely based on pilot symbols, but also on data symbols. This approach is an instance of a \emph{joint processing} of data and pilot symbols in contrast to the separate processing in conventional receiver design. Obviously, this joint processing results in an increased receiver complexity. To evaluate the payoff for the increased receiver complexity, it is important to study the possible performance gain that can be achieved by a joint processing, e.g., in form of an iterative code-aided channel estimation and decoding based receiver, in comparison to a \emph{separate processing} as it is performed in conventional synchronized detection based receivers, where the channel estimation is solely based on pilot symbols. 

Therefore, in the present work we evaluate the performance of a joint processing in comparison to synchronized detection with a solely pilot based channel estimation based on the achievable rate. Regarding the channel statistics, we assume a stationary Rayleigh flat-fading channel as it is usually applied to model the fading in a mobile environment without a line of sight component. Furthermore, we assume that the power spectral density (PSD) of the channel fading process is compactly supported, and that the fading process is \emph{non-regular} \cite{Doob90}, which is reasonable as the maximum Doppler frequency of typical fading channels is small in comparison to the inverse of the symbol duration. Furthermore, we assume that the receiver is aware of the law of the channel, while neither the transmitter nor the receiver knows the realization of the channel fading process.

There has been a variety of publications studying the achievable rate with pilot symbols, see, e.g., \cite{Med00,HassibiHochwald03,Baltersee2001,Baltersee2001b,Furrer2007,LapSha02,ZhTs02}. Many of these works discuss the achievable rate under the assumption that a channel estimate is acquired based on pilot symbols which is then used for coherent detection, i.e., separate processing. Some of these works consider block-fading, \cite{HassibiHochwald03,Furrer2007}, and \cite{ZhTs02}, while \cite{Baltersee2001} and \cite{Baltersee2001b} specifically discuss the case of stationary fading. For the case of a stationary single-input single-output Rayleigh flat-fading channel, as we study in the present work, tight bounds on the achievable rate with synchronized detection with a solely pilot based channel estimation, i.e., separate processing, have been given in \cite{Baltersee2001}. In contrast, for the case of a joint processing there is not much knowledge on the achievable rate. Very recently, in \cite{JindalLozanoMarzettaISIT09} the value of joint processing of pilot and data symbols has been studied in the context of a block-fading channel. To the best of our knowledge, there are no results concerning the gain of joint processing of pilot and data symbols for the case of stationary fading channels. Thus, in the present work, we study the achievable rate with a joint processing of pilot and data symbols. We identify the nature of the possible gain of a joint processing of pilot and data symbols in comparison to a separate processing. Furthermore, we derive a lower bound on the achievable rate with joint processing of pilot and data symbols, which, thus, can be seen as an extension of the work given in \cite{JindalLozanoMarzettaISIT09} to the case of stationary Rayleigh flat-fading. In addition, we compare the given lower bound on the achievable rate with joint processing of pilot and data symbols to bounds on the achievable rate with separate processing given in \cite{Baltersee2001} and to bounds on the achievable rate with i.i.d.\ zero-mean proper Gaussian input symbols given in \cite{Doer08ISITA}, i.e., without the assumption on pilot symbols inserted into the transmit sequence.

The rest of the paper is organized as follows. In Section~\ref{Sect_SysModel} the system model is introduced. Subsequently, in Section~\ref{SectionNature} we discuss the nature of the gain by a joint processing of pilot and data symbols, i.e., we discuss which information is discarded in case of a separate processing. Furthermore, existing bounds on the achievable rate with separate processing are briefly recalled. Afterwards, in Section~\ref{Sect_JointProcessing} a new lower bound on the achievable rate with a joint processing of pilot and data symbols is derived, before it is numerically evaluated and compared to the achievable rate with separate processing and to the achievable rate with i.i.d.\ zero-mean proper Gaussian inputs in Section~\ref{Sect_NumericalEvaluation}. Finally, Section~\ref{Sect_Summary} concludes the paper with a brief summary.

\section{System Model}\label{Sect_SysModel}
We consider a discrete-time zero-mean jointly proper Gaussian flat-fading channel with the following input-output relation
\begin{align}
\mathbf{y}&=\mathbf{H}\mathbf{x}+\mathbf{n}=\mathbf{X}\mathbf{h}+\mathbf{n}\label{ChannelModel1}
\end{align}
with the diagonal matrices $\mathbf{H}=\textrm{diag}(\mathbf{h})$ and $\mathbf{X}=\textrm{diag}(\mathbf{x})$. Here the $\textrm{diag}(\cdot)$ operator generates a diagonal matrix whose diagonal elements are given by the argument vector. The vector $\mathbf{y}=\left[y_{1},\hdots,y_{N}\right]^{T}$ contains the channel output symbols in temporal order. Analogous, $\mathbf{x}=\left[x_{1},\hdots,x_{N}\right]^{T}$, $\mathbf{n}=\left[n_{1},\hdots,n_{N}\right]^{T}$, and $\mathbf{h}=\left[h_{1},\hdots,h_{N}\right]^{T}$ contain the channel input symbols, the additive noise samples and the channel fading weights. All vectors are of length $N$.

The samples of the additive noise process are assumed to be i.i.d.\ zero-mean jointly proper Gaussian with variance $\sigma_{n}^{2}$ and, thus, $\mathbf{R}_{n}=\mathrm{E}\left[\mathbf{n}\mathbf{n}^{H}\right]=\sigma_{n}^{2}\mathbf{I}_{N}$, with $\mathbf{I}_{N}$ being the identity matrix of size $N\times N$.

The channel fading process is zero-mean jointly proper Gaussian with the temporal correlation characterized by
\begin{align}
r_{h}(l)&=\mathrm{E}[h_{k+l}\cdot h_{k}^{*}].\label{DefAutoCorrPrim}
\end{align}
Its variance is given by $r_{h}(0)=\sigma_{h}^{2}$. For mathematical reasons we assume that the autocorrelation function $r_{h}(l)$ is absolutely summable, i.e., 
\begin{align}
\sum_{l=-\infty}^{\infty}|r_{h}(l)|&<\infty. \label{ReqAbsSum}
\end{align}
The PSD of the channel fading process is defined as
\begin{align}
S_{h}(f)&=\sum_{m=-\infty}^{\infty}r_{h}(m)e^{-j2\pi mf}, \qquad |f|\le 0.5.\label{DefPSDH}
\end{align}
We assume that the PSD exists, which for a jointly proper Gaussian fading process implies ergodicity. Furthermore, we assume the PSD to be compactly supported within the interval $[-f_{d},f_{d}]$ with $f_{d}$ being the maximum Doppler shift and $0<f_{d}< 0.5$. This means that $S_{h}(f)=0$ for $f\notin [-f_{d},f_{d}]$. The assumption of a PSD with limited support is motivated by the fact that the velocity of the transmitter, the receiver, and of objects in the environment is limited. To ensure ergodicity, we exclude the case $f_{d}=0$. In matrix-vector notation, the temporal correlation is expressed by the autocorrelation matrix $\mathbf{R}_{h}$ given by
\begin{align}
\mathbf{R}_{h}&=\mathrm{E}\left[\mathbf{h}\mathbf{h}^{H}\right].
\end{align}

For the following derivation we introduce the subvectors $\mathbf{x}_{D}$ containing all data symbols of $\mathbf{x}$ and the vector $\mathbf{x}_{P}$ containing all pilot symbols of $\mathbf{x}$. Correspondingly, we define the vectors $\mathbf{h}_{D}$, $\mathbf{h}_{P}$, $\mathbf{y}_{D}$, $\mathbf{y}_{P}$, $\mathbf{n}_{D}$, and $\mathbf{n}_{P}$. 

The transmit symbol sequence consists of data symbols with a maximal average power $\sigma_{x}^{2}$, i.e., 
\begin{align}
\frac{1}{N_{D}}\mathrm{E}\left[\mathbf{x}_{D}^{H}\mathbf{x}_{D}\right]&\le\sigma_{x}^{2}\label{AveragePowerConstraintJoint}
\end{align}
with $N_{D}$ being the length of the vector $\mathbf{x}_{D}$, and periodically inserted pilot symbols with a fixed transmit power $\sigma_{x}^{2}$. Each $L$-th symbol is a pilot symbol. We assume that the pilot spacing is chosen such that the channel fading process is sampled at least with Nyquist rate, i.e., 
\begin{align}
L&< \frac{1}{2f_{d}}.\label{NyquistCongChannelSamp}
\end{align}

The processes $\{x_{k}\}$, $\{h_{k}\}$ and $\{n_{k}\}$ are assumed to be mutually independent.

Based on the preceding definitions the average SNR $\rho$ is given by
\begin{align}
\rho&=\frac{\sigma_{x}^{2}\sigma_{h}^{2}}{\sigma_{n}^{2}}.
\end{align}

\section{The Nature of the Gain by Joint Processing of Data and Pilot Symbols}\label{SectionNature}
Before we quantitatively discuss the value of a joint processing of data and pilot symbols, we discuss the nature of the possible gain of such a joint processing in comparison to a separate processing of data and pilot symbols. The mutual information between the transmitter and the receiver is given by $\mathcal{I}(\mathbf{x}_{D};\mathbf{y}_{D},\mathbf{y}_{P},\mathbf{x}_{P})$. As the pilot symbols are known to the receiver, the pilot symbol vector $\mathbf{x}_{P}$ is found at the RHS of the semicolon. We separate $\mathcal{I}(\mathbf{x}_{D};\mathbf{y}_{D},\mathbf{y}_{P},\mathbf{x}_{P})$ as follows
\begin{align}
\mathcal{I}(\mathbf{x}_{D};\mathbf{y}_{D},\mathbf{y}_{P},\mathbf{x}_{P})&\stackrel{(a)}{=}\mathcal{I}(\mathbf{x}_{D};\mathbf{y}_{D}|\mathbf{y}_{P},\mathbf{x}_{P})+\mathcal{I}(\mathbf{x}_{D};\mathbf{y}_{P}|\mathbf{x}_{P})+\mathcal{I}(\mathbf{x}_{D};\mathbf{x}_{P})\nonumber\\
&\stackrel{(b)}{=}\mathcal{I}(\mathbf{x}_{D};\mathbf{y}_{D}|\mathbf{y}_{P},\mathbf{x}_{P})\label{EqNature1}
\end{align}
where (a) follows from the chain rule for mutual information and (b) holds due to the independency of the data and pilot symbols. The question is, which portion of $\mathcal{I}(\mathbf{x}_{D};\mathbf{y}_{D}|\mathbf{y}_{P},\mathbf{x}_{P})$ can be achieved by synchronized detection with a solely pilot based channel estimation, i.e., with separate processing.

\subsection{Separate Processing}
The receiver has to find the most likely data sequence $\mathbf{x}_{D}$ based on the observation $\mathbf{y}$ while knowing the pilots $\mathbf{x}_{P}$, i.e.,
\begin{align}
\hat{\mathbf{x}}_{D}
&=\arg\max_{\mathbf{x}_{D}\in \mathcal{C}_{D}}p(\mathbf{y}|\mathbf{x})=\arg\max_{\mathbf{x}_{D}\in \mathcal{C}_{D}}p(\mathbf{y}_{D}|\mathbf{x}_{D},\mathbf{y}_{P},\mathbf{x}_{P})\label{MLDetectionMetricFirst}
\end{align}
with the set $\mathcal{C}_{D}$ containing all possible data sequences $\mathbf{x}_{D}$. It can be shown that the probability density function (PDF) $p(\mathbf{y}_{D}|\mathbf{x}_{D},\mathbf{y}_{P},\mathbf{x}_{P})$ is proper Gaussian and, thus, is completely described by the conditional mean and covariance
\begin{align}
\mathrm{E}\left[\mathbf{y}_{D}|\mathbf{x}_{D},\mathbf{y}_{P},\mathbf{x}_{P}\right]
&=\mathbf{X}_{D}\mathrm{E}\left[\mathbf{h}_{D}|\mathbf{y}_{P},\mathbf{x}_{P}\right]=\mathbf{X}_{D}\hat{\mathbf{h}}_{\textrm{pil},D}\label{Mean_Cond_onlyD}\\
\mathrm{cov}[\mathbf{y}_{D}|\mathbf{x}_{D},\mathbf{y}_{P},\mathbf{x}_{P}]
&=\mathbf{X}_{D}\mathbf{R}_{e_{\textrm{pil}},D}\mathbf{X}_{D}^{H}+\sigma_{n}^{2}\mathbf{I}_{N_{D}}\label{Cov_Cond_onlyD}
\end{align}
where $\mathbf{X}_{D}=\mathrm{diag}(\mathbf{x}_{D})$ and $\mathbf{I}_{N_{D}}$ is an identity matrix of size $N_{D}\times N_{D}$. The vector $\hat{\mathbf{h}}_{\textrm{pil},D}$ is an MMSE channel estimate at the data symbol time instances based on the pilot symbols, which is denoted by the index $pil$. Furthermore, the corresponding channel estimation error 
\begin{align}
\mathbf{e}_{\mathrm{pil},D}&=\mathbf{h}_{D}-\hat{\mathbf{h}}_{\mathrm{pil},D}\label{PilotEstError_new}
\end{align}
is zero-mean proper Gaussian and 
\begin{align}
\mathbf{R}_{e_{\textrm{pil}},D}&=\mathrm{E}\left[\mathbf{e}_{\mathrm{pil},D}\mathbf{e}_{\mathrm{pil},D}^{H}|\mathbf{x}_{P}\right]
\end{align} 
is its correlation matrix, which is independent of $\mathbf{y}_{P}$ due to the principle of orthogonality. 

Based on (\ref{Mean_Cond_onlyD}) and (\ref{Cov_Cond_onlyD}) conditioning of $\mathbf{y}_{D}$ on $\mathbf{x}_{D},\mathbf{y}_{P},\mathbf{x}_{P}$ is equivalent to conditioning on $\mathbf{x}_{D},\hat{\mathbf{h}}_{\mathrm{pil},D},\mathbf{x}_{P}$, i.e.,
\begin{align}
p(\mathbf{y}_{D}|\mathbf{x}_{D},\mathbf{y}_{P},\mathbf{x}_{P})&=p(\mathbf{y}_{D}|\mathbf{x}_{D},\hat{\mathbf{h}}_{\mathrm{pil},D},\mathbf{x}_{P})\label{PDFSubstEquality}
\end{align}
as all information on $\mathbf{h}_{D}$ delivered by $\mathbf{y}_{P}$ is contained in $\hat{\mathbf{h}}_{\textrm{pil},D}$ while conditioning on $\mathbf{x}_{P}$. Thus, (\ref{MLDetectionMetricFirst}) can be written as
\begin{align}
\hat{\mathbf{x}}_{D}&=\arg\max_{\mathbf{x}_{D}\in \mathcal{C}_{D}}p(\mathbf{y}_{D}|\mathbf{x}_{D},\hat{\mathbf{h}}_{\mathrm{pil},D},\mathbf{x}_{P})=\arg\max_{\mathbf{x}_{D}\in \mathcal{C}_{D}}p(\mathbf{y}|\mathbf{x}_{D},\hat{\mathbf{h}}_{\mathrm{pil}},\mathbf{x}_{P}).\label{MLDetectMetricCompete}
\end{align} 
For ease of notation in the following we will use the metric on the RHS of (\ref{MLDetectMetricCompete}) where $\hat{\mathbf{h}}_{\textrm{pil}}$ corresponds to $\hat{\mathbf{h}}_{\textrm{pil},D}$ but also contains channel estimates at the pilot symbol time instances, i.e., 
\begin{align}
\hat{\mathbf{h}}_{\textrm{pil}}&=\mathrm{E}\left[\mathbf{h}|\mathbf{y}_{P},\mathbf{x}_{P}\right].
\end{align}
Based on $\hat{\mathbf{h}}_{\textrm{pil}}$, (\ref{ChannelModel1}) can be expressed by\looseness-1
\begin{align}
\mathbf{y}&=\mathbf{X}(\hat{\mathbf{h}}_{\textrm{pil}}+\mathbf{e}_{\textrm{pil}})+\mathbf{n}\label{IORelationEstBased}
\end{align}
where $\mathbf{e}_{\textrm{pil}}$ is the estimation error including the pilot symbol time instances. As the channel estimation is an interpolation, the error process is not white but temporally correlated, i.e.,
\begin{align}
\mathbf{R}_{e_{\textrm{pil}}}&=\mathrm{E}\left[\mathbf{e}_{\textrm{pil}}\mathbf{e}_{\textrm{pil}}^{H}|\mathbf{x}_{P}\right]\label{ErrorCorr_Pil}
\end{align}
is not diagonal, cf. (\ref{PSE4}). As the estimation error process is zero-mean proper Gaussian, the PDF in (\ref{MLDetectMetricCompete}) is given by\looseness-1
\begin{align}
p(\mathbf{y}|\mathbf{x}_{D},\hat{\mathbf{h}}_{\textrm{pil}},\mathbf{x}_{P})&=\mathcal{CN}\left(\mathbf{X}\hat{\mathbf{h}}_{\textrm{pil}},\mathbf{X}\mathbf{R}_{e_{\textrm{pil}}}\mathbf{X}^{H}+\sigma_{n}^{2}\mathbf{I}_{N}\right)\label{Conhathydensity}
\end{align}
where $\mathcal{CN}(\boldsymbol{\mu},\mathbf{C})$ denotes a proper Gaussian PDF with mean $\boldsymbol{\mu}$ and covariance $\mathbf{C}$ and where $\mathbf{I}_{N}$ is the $N\times N$ identity matrix.\footnote{Note that for the case of data transmission only (\ref{Conhathydensity}) becomes $p(\mathbf{y}|\mathbf{x}_{D})= \mathcal{CN}(\mathbf{0},\mathbf{X}\mathbf{R}_{h}\mathbf{X}^{H}+\sigma_{n}^{2}\mathbf{I}_{N})$ as in this case $\hat{\mathbf{h}}_{\textrm{pil}}=\mathbf{0}$ and $\mathbf{R}_{e_{\textrm{pil}}}=\mathbf{R}_{h}$.}\looseness-1

Corresponding to (\ref{PDFSubstEquality}), we can also rewrite $p(\mathbf{y}_{D}|\mathbf{y}_{P},\mathbf{x}_{P})$ as follows
\begin{align}
p(\mathbf{y}_{D}|\mathbf{y}_{P},\mathbf{x}_{P})&=\int p(\mathbf{y}_{D}|\mathbf{x}_{D},\mathbf{y}_{P},\mathbf{x}_{P})p(\mathbf{x}_{D}|\mathbf{y}_{P},\mathbf{x}_{P})d\mathbf{x}_{D}\nonumber\\
&\stackrel{(a)}{=}\int p(\mathbf{y}_{D}|\mathbf{x}_{D},\hat{\mathbf{h}}_{\mathrm{pil},D},\mathbf{x}_{P})p(\mathbf{x}_{D})d\mathbf{x}_{D}\nonumber\\
&=p(\mathbf{y}_{D}|\hat{\mathbf{h}}_{\mathrm{pil},D},\mathbf{x}_{P})\label{Subshathpil_2_Joint}
\end{align}
where for (a) we have used (\ref{PDFSubstEquality}) and the independency of $\mathbf{x}_{D}$ of $\mathbf{x}_{P}$ and $\mathbf{y}_{P}$.

Based on (\ref{PDFSubstEquality}) and (\ref{Subshathpil_2_Joint}), we can also rewrite (\ref{EqNature1}) as
\begin{align}
&\mathcal{I}(\mathbf{x}_{D};\mathbf{y}_{D}|\mathbf{y}_{P},\mathbf{x}_{P})=\mathcal{I}(\mathbf{x}_{D};\mathbf{y}_{D}|\hat{\mathbf{h}}_{\textrm{pil}},\mathbf{x}_{P})\stackrel{(a)}{=}\mathcal{I}(\mathbf{x}_{D};\mathbf{y}_{D}|\hat{\mathbf{h}}_{\textrm{pil}})\label{RewritwMutInfoCond}
\end{align}
and where (a) holds as the pilot symbols are deterministic.

However, typical channel decoders like a Viterbi decoder are not able to exploit the temporal correlation of the channel estimation error. Therefore, the decoder performs mismatched decoding based on the assumption that the estimation error process is white, i.e., $p(\mathbf{y}|\mathbf{x}_{D},\hat{\mathbf{h}}_{\textrm{pil}},\mathbf{x}_{P})$ is approximated by
\begin{align}
p(\mathbf{y}|\mathbf{x}_{D},\hat{\mathbf{h}}_{\textrm{pil}},\mathbf{x}_{P})&\approx\mathcal{CN}\left(\mathbf{X}\hat{\mathbf{h}}_{\textrm{pil}},\sigma_{e_{\textrm{pil}}}^{2}\mathbf{X}\mathbf{X}^{H}+\sigma_{n}^{2}\mathbf{I}_{N}\right).\label{DetMetricPDFApprox}
\end{align}

As it is assumed that the channel is at least sampled with Nyquist frequency, see (\ref{NyquistCongChannelSamp}), for an infinite block length $N\rightarrow \infty$ the channel estimation error variance $\sigma_{e_{\textrm{pil}}}^{2}$ is independent of the symbol time instant \cite{Baltersee2001} and is given by
\begin{align}
\sigma_{e_{\textrm{pil}}}^{2}&=\int_{f=-\frac{1}{2}}^{\frac{1}{2}}S_{e_{\textrm{pil}}}(f)df=\int_{f=-\frac{1}{2}}^{\frac{1}{2}}\frac{S_{h}(f)}{\frac{\rho}{L}\frac{S_{h}(f)}{\sigma_{h}^{2}}+1}df
\end{align}
where $S_{e_{\textrm{pil}}}(f)$ is the PSD of the channel estimation error process in case the channel estimation is solely based on pilot symbols, which is given in (\ref{PSE4_App}) in Appendix~\ref{SectionPilBasedErrorSpec}. Hence, the variance of the channel estimation process, i.e., the entries of $\hat{\mathbf{h}}_{\textrm{pil}}$, is given by $\sigma_{h}^{2}-\sigma_{e_{\textrm{pil}}}^{2}$, which follows from the principle of orthogonality in LMMSE estimation.

As the information contained in the temporal correlation of the channel estimation error is not retrieved by synchronized detection with a solely pilot based channel estimation, the mutual information in this case corresponds to the sum of the mutual information for each individual data symbol time instant. As, obviously, by this separate processing information is discarded, the following inequality for the achievable rate holds:
\begin{align}
\lim_{N\rightarrow\infty}\frac{1}{N}\mathcal{I}(\mathbf{x}_{D};\mathbf{y}_{D}|\hat{\mathbf{h}}_{\textrm{pil}})&=\mathcal{I}'(\mathbf{x}_{D};\mathbf{y}_{D}|\hat{\mathbf{h}}_{\textrm{pil}})\nonumber\\
&\ge \frac{L-1}{L}\mathcal{I}(x_{D_{k}};y_{D_{k}}|\hat{\mathbf{h}}_{\textrm{pil}})\nonumber\\
&=\frac{L-1}{L}\mathcal{I}(x_{D_{k}};y_{D_{k}}|\hat{h}_{\textrm{pil},D_{k}})
=\mathcal{R}_{\textrm{sep}}\label{InequalDiscardTemp}
\end{align}
where $\mathcal{I}'$ denotes the mutual information rate and the index $D_{k}$ refers to an arbitrarily chosen data symbol, i.e., $x_{D_{k}}=\left[\mathbf{x}_{D}\right]_{k}$. Furthermore, $\hat{h}_{\textrm{pil},D_{k}}$ is the solely pilot based channel estimate at the data symbol time instant $D_{k}$. The pre-factor $(L-1)/L$ arises from the fact that each $L$-th symbol is a pilot symbol. In the following, we denote the achievable rate with separate processing by $\mathcal{R}_{\textrm{sep}}$.

As the LHS of (\ref{InequalDiscardTemp}) is the mutual information of the channel and as the RHS of (\ref{InequalDiscardTemp}) is the mutual information achievable with synchronized detection with a metric corresponding to (\ref{DetMetricPDFApprox}) and a solely pilot based channel estimation, i.e., a separate processing, the difference of both terms upper bounds the possible gain due to joint processing of data and pilot symbols. Obviously, the additional information that can be gained by a joint processing in contrast to the separate processing is contained in the temporal correlation of the channel estimation error process.

Regarding synchronized detection in combination with a solely pilot based channel estimation, i.e., the separate processing approach, in \cite{Baltersee2001} bounds on the achievable rate have been given, which for zero-mean proper Gaussian data symbols become
\begin{align}
\mathcal{R}_{\textrm{sep}}\ge\mathcal{R}_{L,\textrm{sep}}&=\frac{L-1}{L}\mathrm{E}_{\hat{h}_{\textrm{pil},D_{k}}}\left[\log\left(1+\frac{\sigma_{x}^{2}|\hat{h}_{\textrm{pil},D_{k}}|^{2}}{\sigma_{e_{\textrm{pil}}}^{2}\sigma_{x}^{2}+\sigma_{n}^{2}}\right)\right]\nonumber\\
&=\frac{L-1}{L}\int_{z=0}^{\infty}\log\left(1+\rho\frac{1-\frac{\sigma_{e_{\textrm{pil}}}^{2}}{\sigma_{h}^{2}}}{1+\rho\frac{\sigma_{e_{\textrm{pil}}}^{2}}{\sigma_{h}^{2}}} z\right)e^{-z}dz\label{LowerBoundSD}\\
\mathcal{R}_{\textrm{sep}}\le\mathcal{R}_{U,\textrm{sep}}&=\mathcal{R}_{L,\textrm{sep}}+\frac{L-1}{L}\mathrm{E}_{x_{D_{k}}}\left[\log\left(\frac{\sigma_{x}^{2}\sigma_{e_{\textrm{pil}}}^{2}+\sigma_{n}^{2}}{|x_{D_{k}}|^{2}\sigma_{e_{\textrm{pil}}}^{2}+\sigma_{n}^{2}}\right)\right]\nonumber\\
&=\mathcal{R}_{L,\textrm{sep}}+\frac{L-1}{L}\Bigg(\log\left(1+\rho\frac{\sigma_{e_{\textrm{pil}}}^{2}}{\sigma_{h}^{2}}\right)-\int_{z=0}^{\infty}\log\left(1+\rho\frac{\sigma_{e_{\textrm{pil}}}^{2}}{\sigma_{h}^{2}} z\right)e^{-z}dz\Bigg).\label{UpperBoundSD}
\end{align}
Based on the lower bound in (\ref{LowerBoundSD}) it can easily be seen that the achievable rate is decreased in comparison to perfect channel knowledge by two factors. First, symbol time instances that are used for pilot symbols are lost for data symbols leading to the pre-log factor $\frac{L-1}{L}$, and secondly, the average SNR is decreased by the factor $\left(1-\frac{\sigma_{e_{\textrm{pil}}}^{2}}{\sigma_{h}^{2}}\right)/\left(1+\rho\frac{\sigma_{e_{\textrm{pil}}}^{2}}{\sigma_{h}^{2}}\right)$ due to the channel estimation error variance. The additional term in the upper bound in (\ref{UpperBoundSD}) arises from the fact that the effective noise, i.e., $e_{\textrm{pil},D_{k}}x_{D_{k}}+n_{D_{k}}$, is non-Gaussian. Here $e_{D_{k}}$ is the estimation error at the data symbol time instant $D_{k}$, i.e., $e_{D_{k}}=\left[\mathbf{e}_{\textrm{pil},D}\right]_{k}$.

\section{Joint Processing of Data and Pilot Symbols}\label{Sect_JointProcessing}
Now, we give a new lower bound on the achievable rate for a joint processing of data and pilot symbols. The following approach can be seen as an extension of the work in \cite{JindalLozanoMarzettaISIT09} for the case of a block-fading channel to the stationary Rayleigh flat-fading scenario discussed in the present work. Therefore, analogously to \cite{JindalLozanoMarzettaISIT09} we decompose and lower-bound the mutual information between the transmitter and the receiver $\mathcal{I}(\mathbf{x}_{D};\mathbf{y}_{D},\mathbf{y}_{P},\mathbf{x}_{P})$ as follows
\begin{align}
\mathcal{I}(\mathbf{x}_{D};\mathbf{y}_{D},\mathbf{y}_{P},\mathbf{x}_{P})&\stackrel{(a)}{=}\mathcal{I}(\mathbf{x}_{D};\mathbf{y}_{D},\mathbf{y}_{P},\mathbf{x}_{P},\mathbf{h})-\mathcal{I}(\mathbf{x}_{D};\mathbf{h}|\mathbf{y}_{D},\mathbf{y}_{P},\mathbf{x}_{P})\nonumber\\
&\stackrel{(b)}{=}\mathcal{I}(\mathbf{x}_{D};\mathbf{y}_{D},\mathbf{h})-h(\mathbf{h}|\mathbf{y}_{D},\mathbf{y}_{P},\mathbf{x}_{P})+h(\mathbf{h}|\mathbf{x}_{D},\mathbf{y}_{D},\mathbf{y}_{P},\mathbf{x}_{P})\nonumber\\
&\stackrel{(c)}{\ge}\mathcal{I}(\mathbf{x}_{D};\mathbf{y}_{D},\mathbf{h})-h(\mathbf{h}|\mathbf{y}_{P},\mathbf{x}_{P})+h(\mathbf{h}|\mathbf{x}_{D},\mathbf{y}_{D},\mathbf{y}_{P},\mathbf{x}_{P})\label{LowerBoundJoint1}
\end{align}
where (a) follows from the chain rule for mutual information. For the first term in (b)  we have used the fact that due to the knowledge on $\mathbf{h}$, the knowledge on $\mathbf{y}_{P}$ and $\mathbf{x}_{P}$ does not increase the mutual information between $\mathbf{x}_{D}$ and $\mathbf{y}_{D}$. Finally, (c) is due to the fact that conditioning reduces entropy. Note, the first term on the RHS of (\ref{LowerBoundJoint1}) is the mutual information in case of perfect channel knowledge. 

In the following we deviate from the derivation given in \cite{JindalLozanoMarzettaISIT09}. Now, we calculate both differential entropy terms at the RHS of (\ref{LowerBoundJoint1}). Therefore, we rewrite the RHS of (\ref{LowerBoundJoint1}) as follows
\begin{align}
\mathcal{I}(\mathbf{x}_{D};\mathbf{y}_{D},\mathbf{y}_{P},\mathbf{x}_{P})&\ge \mathcal{I}(\mathbf{x}_{D};\mathbf{y}_{D},\mathbf{h})-h(\mathbf{h}|\mathbf{y}_{P},\mathbf{x}_{P})+h(\mathbf{h}|\mathbf{x}_{D},\mathbf{y}_{D},\mathbf{y}_{P},\mathbf{x}_{P})\nonumber\\
&\stackrel{(a)}{=}\mathcal{I}(\mathbf{x}_{D};\mathbf{y}_{D},\mathbf{h})-h(\mathbf{h}|\hat{\mathbf{h}}_{\textrm{pil}},\mathbf{x}_{P})+h(\mathbf{h}|\hat{\mathbf{h}}_{\textrm{joint}},\mathbf{x}_{D},\mathbf{x}_{P})\nonumber\\
&\stackrel{(b)}{=}\mathcal{I}(\mathbf{x}_{D};\mathbf{y}_{D},\mathbf{h})-h(\hat{\mathbf{h}}_{\textrm{pil}}+\mathbf{e}_{\textrm{pil}}|\hat{\mathbf{h}}_{\textrm{pil}},\mathbf{x}_{P})+h(\hat{\mathbf{h}}_{\textrm{joint}}+\mathbf{e}_{\textrm{joint}}|\hat{\mathbf{h}}_{\textrm{joint}},\mathbf{x}_{D},\mathbf{x}_{P})\nonumber\\
&\stackrel{(c)}{=}\mathcal{I}(\mathbf{x}_{D};\mathbf{y}_{D},\mathbf{h})-h(\mathbf{e}_{\textrm{pil}}|\mathbf{x}_{P})+h(\mathbf{e}_{\textrm{joint}}|\mathbf{x}_{D},\mathbf{x}_{P})\nonumber\\
&\stackrel{(d)}{=}\mathcal{I}(\mathbf{x}_{D};\mathbf{y}_{D},\mathbf{h})-\mathrm{E}_{\mathbf{x}_{P}}\left[\log\det\left(\pi e\mathbf{R}_{e_{\textrm{pil}}}\right)\right]+\mathrm{E}_{\mathbf{x}_{P},\mathbf{x}_{D}}\left[\log\det\left(\pi e\mathbf{R}_{e_{\textrm{joint}}}\right)\right]\nonumber\\
&\stackrel{(e)}{=}\mathcal{I}(\mathbf{x}_{D};\mathbf{y}_{D},\mathbf{h})-\log\det\left(\mathbf{R}_{e_{\textrm{pil}}}\right)+\mathrm{E}_{\mathbf{x}_{D}}\left[\log\det\left(\mathbf{R}_{e_{\textrm{joint}}}\right)\right]\label{LowerBoundJoint2}
\end{align}
where for the second term in (a) we have substituted the condition on $\mathbf{y}_{P}$ by $\hat{\mathbf{h}}_{\textrm{pil}}$, which is possible as the estimate $\hat{\mathbf{h}}_{\textrm{pil}}$ contains the same information on $\mathbf{h}$ as $\mathbf{y}_{P}$ while conditioning on $\mathbf{x}_{P}$. Corresponding to the solely pilot based channel estimate $\hat{\mathbf{h}}_{\textrm{pil}}$, based on $\mathbf{x}_{D}$, $\mathbf{x}_{P}$, $\mathbf{y}_{D}$, and $\mathbf{y}_{P}$, we can calculate the estimate $\hat{\mathbf{h}}_{\textrm{joint}}$, which is based on data and pilot symbols. Like $\hat{\mathbf{h}}_{\textrm{pil}}$ this estimate is a MAP estimate, which, due to the jointly Gaussian nature of the problem, is an MMSE estimate, i.e., 
\begin{align}
\hat{\mathbf{h}}_{\textrm{joint}}&=\mathrm{E}\left[\mathbf{h}|\mathbf{y}_{P},\mathbf{x}_{P},\mathbf{y}_{D},\mathbf{x}_{D}\right].
\end{align}
Thus, for (a) we have substituted the conditioning on $\mathbf{y}_{D}$ and $\mathbf{y}_{P}$ by conditioning on $\hat{\mathbf{h}}_{\textrm{joint}}$ in the third term, as $\hat{\mathbf{h}}_{\textrm{joint}}$ contains all information on $\mathbf{h}$ that is contained in $\mathbf{y}_{D}$ and $\mathbf{y}_{P}$ while $\mathbf{x}_{D}$ and $\mathbf{x}_{P}$ are known. For equality (b) we have used for the second term that $\mathbf{h}$ can be expressed as a sum of its estimate $\hat{\mathbf{h}}_{\textrm{pil}}$ and the estimation error $\mathbf{e}_{\textrm{pil}}$, cf. (\ref{IORelationEstBased}). Analogously, for the third term we used the separation of $\mathbf{h}$ into the estimate $\hat{\mathbf{h}}_{\textrm{joint}}$ and the corresponding estimation error $\mathbf{e}_{\textrm{joint}}$, i.e.,
\begin{align}
\mathbf{e}_{\textrm{joint}}&=\mathbf{h}-\hat{\mathbf{h}}_{\textrm{joint}}.
\end{align}
Equality (c) is due to the fact that the addition of a constant does not change differential entropy and that the estimation error $\mathbf{e}_{\textrm{pil}}$ is independent of the estimate $\hat{\mathbf{h}}_{\textrm{pil}}$ and analogously $\mathbf{e}_{\textrm{joint}}$, which depends on $\mathbf{x}_{P}$ and $\mathbf{x}_{D}$, is independent of $\hat{\mathbf{h}}_{\textrm{joint}}$ due to the orthogonality principle in LMMSE estimation. Finally, (d) follows from the fact that the estimation error processes are zero-mean jointly proper Gaussian. Here the error correlation matrices are given by (\ref{ErrorCorr_Pil}) and by
\begin{align}
\mathbf{R}_{e_{\textrm{joint}}}&=\mathrm{E}\left[\mathbf{e}_{\textrm{joint}}\mathbf{e}_{\textrm{joint}}^{H}|\mathbf{x}_{D},\mathbf{x}_{P}\right].\label{CorrJoint_gen}
\end{align}

For (e) we have used that the pilot symbols are deterministic. Therefore, the expectation over $\mathbf{x}_{P}$ in the second and third term can be removed. However, the channel estimation error $\mathbf{e}_{\textrm{joint}}$ depends on the distribution of the data symbols $\mathbf{x}_{D}$. Concerning the third term on the RHS of (\ref{LowerBoundJoint2}), it can be shown that the differential entropy rate $h'(\mathbf{e}_{\textrm{joint}}|\mathbf{x}_{D},\mathbf{x}_{P})$, i.e.,
\begin{align}
h'(\mathbf{e}_{\textrm{joint}}|\mathbf{x}_{D},\mathbf{x}_{P})&=\lim_{N\rightarrow \infty}\frac{1}{N}h(\mathbf{e}_{\textrm{joint}}|\mathbf{x}_{D},\mathbf{x}_{P})\label{DiffEntropyJointErrorRate_Joint}
\end{align}
is minimized for a given average transmit power $\sigma_{x}^{2}$ if the data symbols are constant modulus (CM) symbols with power $\sigma_{x}^{2}$, see Appendix~\ref{ProofLowerrBoundCM}. Within this proof the restriction to an absolutely summable autocorrelation function $r_{h}(l)$, see (\ref{ReqAbsSum}), is required.

Thus, based on (\ref{LowerBoundJoint2}) a lower bound for the achievable rate with joint processing of data and pilot symbols is given by
\begin{align}
\mathcal{I}'(\mathbf{x}_{D};\mathbf{y}_{D},\mathbf{y}_{P},\mathbf{x}_{P})&=\lim_{N\rightarrow \infty}\frac{1}{N}\mathcal{I}(\mathbf{x}_{D};\mathbf{y}_{D},\mathbf{y}_{P},\mathbf{x}_{P})\nonumber\\
&\ge\lim_{N\rightarrow \infty}\frac{1}{N}\left\{\mathcal{I}(\mathbf{x}_{D};\mathbf{y}_{D},\mathbf{h})-\log\det\left(\mathbf{R}_{e_{\textrm{pil}}}\right)+\log\det\left(\mathbf{R}_{e_{\textrm{joint,CM}}}\right)\right\}\nonumber\\
&\stackrel{(a)}{=}\lim_{N\rightarrow \infty}\frac{1}{N}\mathcal{I}(\mathbf{x}_{D};\mathbf{y}_{D},\mathbf{h})-\int_{-\frac{1}{2}}^{\frac{1}{2}}\log\left(\frac{S_{e_{\textrm{pil}}}(f)}{S_{e_{\textrm{joint,CM}}}(f)}\right)df\label{LowerBoundJoint4_gen}
\end{align}
with $\mathbf{R}_{e_{\textrm{joint,CM}}}$ corresponding to (\ref{CorrJoint_gen}), but under the assumption of CM data symbols with transmit power $\sigma_{x}^{2}$.  As $\mathbf{R}_{e_{\textrm{joint,CM}}}$ only depends on the distribution of the magnitude of the data symbols contained in $\mathbf{x}_{D}$, which is constant and deterministic, we can remove the expectation operation with respect to $\mathbf{x}_{D}$. Note that the CM assumption has only been used to lower-bound the third term at the RHS of (\ref{LowerBoundJoint2}), and not the whole expression at the RHS of (\ref{LowerBoundJoint2}). For (a) in (\ref{LowerBoundJoint4_gen}) we have used Szeg\"o's theorem on the asymptotic eigenvalue distribution of Hermitian Toeplitz matrices \cite{Szgo58}. $S_{e_{\textrm{pil}}}(f)$ and $S_{e_{\textrm{joint,CM}}}(f)$ are the PSDs of the channel estimation error processes, on the one hand, if the estimation is solely based on pilot symbols, and on the other hand, if the estimation is based on data and pilot symbols, assuming CM data symbols. They are given by
\begin{align}
S_{e_{\textrm{pil}}}(f)&=\frac{S_{h}(f)}{\frac{\rho}{L} \frac{S_{h}(f)}{\sigma_{h}^{2}}+1}\label{PSE4}\\
S_{e_{\textrm{joint,CM}}}(f)&=\frac{S_{h}(f)}{\rho \frac{S_{h}(f)}{\sigma_{h}^{2}}+1}.\label{PSE5}
\end{align}
The derivation of these PSDs is given in Appendix~\ref{SectionPilBasedErrorSpec}.

However, the application of Szeg\"o's theorem for (a) in (\ref{LowerBoundJoint4_gen}) requires several steps, which we discuss in the following. The limit over the second and the third term on the LHS of (a) in (\ref{LowerBoundJoint4_gen}) can be transformed as follows
\begin{align}
&\lim_{N\rightarrow \infty}\frac{1}{N}\left\{\log\det\left(\mathbf{R}_{e_{\textrm{pil}}}\right)-\log\det\left(\mathbf{R}_{e_{\textrm{joint,CM}}}\right)\right\}\nonumber\\
&\qquad\stackrel{(a)}{=}\lim_{N\rightarrow \infty}\frac{1}{N}\left\{\log\det\left(\mathbf{C}_{e_{\textrm{pil}}}\right)-\log\det\left(\mathbf{C}_{e_{\textrm{joint,CM}}}\right)\right\}\nonumber\\
&\qquad\stackrel{(b)}{=}\lim_{N\rightarrow \infty}\frac{1}{N}\left\{\log\det\left(\mathbf{F}\mathbf{\Lambda}_{e_{\textrm{pil}}}\mathbf{F}^{H}\right)-\log\det\left(\mathbf{F}\mathbf{\Lambda}_{e_{\textrm{joint,CM}}}\mathbf{F}^{H}\right)\right\}\nonumber\\
&\qquad=\lim_{N\rightarrow \infty}\frac{1}{N}\left\{\log\det\left(\mathbf{F}\mathbf{\Lambda}_{e_{\textrm{pil}}}\mathbf{\Lambda}_{e_{\textrm{joint,CM}}}^{-1}\mathbf{F}^{H}\right)\right\}\nonumber\\
&\qquad\stackrel{(c)}{=}\int_{-\frac{1}{2}}^{\frac{1}{2}}\log\left(\frac{S_{e_{\textrm{pil}}}(f)}{S_{e_{\textrm{joint,CM}}}(f)}\right)df\nonumber\\
&\qquad\stackrel{(d)}{=}\int_{-\frac{1}{2}}^{\frac{1}{2}}\log\left(\frac{\rho\frac{S_{h}(f)}{\sigma_{h}^{2}}+1}{\frac{\rho}{L}\frac{S_{h}(f)}{\sigma_{h}^{2}}+1}\right)df\label{Szegoe2}
\end{align}
where for (a) we have substituted the Toeplitz matrices $\mathbf{R}_{e_{\textrm{pil}}}$ and $\mathbf{R}_{e_{\textrm{joint,CM}}}$ by their asymptotic equivalent circulant matrices $\mathbf{C}_{e_{\textrm{pil}}}$ and $\mathbf{C}_{e_{\textrm{joint,CM}}}$, see \cite{Gray_ToeplitzReview}. Furthermore, for (b) we have used the spectral decompositions of the circulant matrices given by
\begin{align}
\mathbf{C}_{e_{\textrm{pil}}}&=\mathbf{F}\mathbf{\Lambda}_{e_{\textrm{pil}}}\mathbf{F}^{H}\\
\mathbf{C}_{e_{\textrm{joint,CM}}}&=\mathbf{F}\mathbf{\Lambda}_{e_{\textrm{joint,CM}}}\mathbf{F}^{H}
\end{align}
where $\mathbf{\Lambda}_{e_{\textrm{pil}}}$ and $\mathbf{\Lambda}_{e_{\textrm{joint,CM}}}$ are diagonal matrices containing the eigenvalues of $\mathbf{C}_{e_{\textrm{pil}}}$ and $\mathbf{C}_{e_{\textrm{joint,CM}}}$, and the matrix $\mathbf{F}$ is a unitary DFT-matrix whose elements are given by
\begin{align}
\left[F\right]_{k,l}&=\frac{1}{\sqrt{N}}e^{j2\pi \frac{(k-1)(l-1)}{N}}.
\end{align}
For (c) in (\ref{Szegoe2}) we have then used Szeg\"o's theorem on the asymptotic eigenvalue distribution of Hermitian Toeplitz matrices \cite{Szgo58}. Therefore, first consider that the matrix $\mathbf{F}\mathbf{\Lambda}_{e_{\textrm{pil}}}\mathbf{\Lambda}_{e_{\textrm{joint,CM}}}^{-1}\mathbf{F}^{H}$ on the LHS of (c) is again a circulant matrix and that there exists an asymptotically equivalent Toeplitz matrix. Furthermore, the eigenvalues of $\mathbf{C}_{e_{\textrm{pil}}}$ are samples of the PSD $S_{e_{\textrm{pil}}}(f)$ and the eigenvalues of $\mathbf{C}_{e_{\textrm{joint,CM}}}$ are samples of the PSD $S_{e_{\textrm{joint,CM}}}(f)$. 
Here we assume a construction of the circulant matrices as described in \cite[(4.32)]{Gray_ToeplitzReview}, see also in Appendix~\ref{ProofLowerrBoundCM} from (\ref{defineContPSD}) to (\ref{DefDFT}). Furthermore, the application of Szeg\"o's theorem requires that the $\log$-function is continuous on the support of the eigenvalues of the matrix $\mathbf{\Lambda}_{e_{\textrm{pil}}}\mathbf{\Lambda}_{e_{\textrm{joint,CM}}}^{-1}$. This means that we have to show that the eigenvalues of $\mathbf{\Lambda}_{e_{\textrm{pil}}}\mathbf{\Lambda}_{e_{\textrm{joint,CM}}}^{-1}$ are bounded away from zero and from infinity. That this is indeed the case will become obvious after introducing $S_{e_{\textrm{pil}}}(f)$ and $S_{e_{\textrm{joint,CM}}}(f)$ given in (\ref{PSE4}) and (\ref{PSE5}) as it has been done in (d). Obviously, the argument of the $\log$ at the RHS of (\ref{Szegoe2}) is larger than zero and smaller than infinity on the interval $f\in [-0.5, 0.5]$. Therefore, the integral on the RHS of (\ref{Szegoe2}) exists, implying that also the LHS of (c) in (\ref{Szegoe2}) is bounded and, thus, that the eigenvalues of $\mathbf{\Lambda}_{e_{\textrm{pil}}}\mathbf{\Lambda}_{e_{\textrm{joint,CM}}}^{-1}$ are bounded away from zero and from infinity. Thus, in conclusion we have shown that Szeg\"o's theorem is applicable and that (a) in (\ref{LowerBoundJoint4_gen}) holds.  

The first term on the RHS of (\ref{LowerBoundJoint4_gen}) is the mutual information rate in case of perfect channel state information, which for an average power constraint is maximized with i.i.d.\ zero-mean proper Gaussian data symbols. Thus, we get the following lower bound on the achievable rate with joint processing
\begin{align}
\mathcal{R}_{L,\textrm{joint}}&=\frac{L-1}{L}C_{\textrm{perf}}(\rho)-\int_{-\frac{1}{2}}^{\frac{1}{2}}\log\left(\frac{\frac{\rho}{\sigma_{h}^{2}}S_{h}(f)+1}{\frac{\rho}{L\sigma_{h}^{2}}S_{h}(f)+1}\right)df\label{LowerBoundJoint4}
\end{align}
where $C_{\textrm{perf}}(\rho)$ corresponds to the coherent capacity with
\begin{align}
C_{\textrm{perf}}(\rho)&=\mathrm{E}_{h_{k}}\left[\log\left(1+\rho\frac{|h_{k}|^{2}}{\sigma_{h}^{2}}\right)\right]=\int_{z=0}^{\infty}\log\left(1+\rho z\right)e^{-z}dz\label{CoherentCapacity}
\end{align}
and the factor $(L-1)/L$ arises as each $L$-th symbol is a pilot symbol.

\subsection{Lower Bound on the Achievable Rate for a Joint Processing of Data and Pilot Symbols and a Fixed Pilot Spacing}
Substituting (\ref{CoherentCapacity}) into (\ref{LowerBoundJoint4}) we have found a lower bound on the achievable rate with joint processing of data and pilot symbols, for a given pilot spacing $L$ and stationary Rayleigh flat-fading.

For the special case of a rectangular PSD\footnote{Note that a rectangular PSD $S_{h}(f)$ corresponds to $r_{h}(l)=\sigma_{h}^{2}\textrm{sinc}(2f_{d}l)$ which is not absolutely summable. However, the rectangular PSD can be arbitrarily closely approximated by a PSD with a raised cosine shape, whose corresponding correlation function is absolutely summable.} of the channel fading process, i.e., 
\begin{align}
S_{h}(f)&=\left\{\begin{array}{ll}
\frac{\sigma_{h}^{2}}{2f_{d}} & \textrm{ for } |f|\le f_{d}\\
0 & \textrm{ otherwise} \end{array}\right.\label{PSDRect}
\end{align}
the lower bound in (\ref{LowerBoundJoint4}) becomes
\begin{align}
\mathcal{R}_{L,\textrm{joint}}\big|_{\textrm{rect.} S_{h}(f)}&=\frac{L-1}{L}\int_{z=0}^{\infty}\log\left(1+\rho z\right)e^{-z}dz-2f_{d}\log\left(\frac{\frac{\rho}{2f_{d}}+1}{\frac{\rho}{L 2f_{d}}+1}\right).\label{PSE6}
\end{align}

\subsection{Lower Bound on the Achievable Rate for a Joint Processing of Data and Pilot Symbols and an Optimal Pilot Spacing} 
Obviously, the lower bound in (\ref{PSE6}) still depends on the pilot spacing $L$. In case the pilot spacing is not fixed, we can further enhance it by calculating the supremum of (\ref{PSE6}) with respect to $L$. In this regard, it has to be considered that the pilot spacing $L$ is an integer value. Furthermore, we have to take into account that the derivation of the lower bound in (\ref{PSE6}) is based on the assumption that the pilot spacing is chosen such that the channel fading process is at least sampled with Nyquist rate, i.e., (\ref{NyquistCongChannelSamp}) has to be fulfilled. In case the pilot spacing $L$ is chosen larger than the Nyquist rate, the estimation error process is no longer stationary, which is required for our derivation. At this point it is also important to remark that periodically inserted pilot symbols do not maximize the achievable rate. For the special case of PSK signaling, it is shown in \cite{Doer08ISIT} that the use of a single pilot symbol, i.e., not periodically inserted pilot symbols, is optimal in the sense that it maximizes the achievable rate. However, in the present work we restrict to the assumption of periodically inserted pilot symbols with a pilot spacing fulfilling (\ref{NyquistCongChannelSamp}), which is customary and reasonable as this enables detection and decoding with manageable complexity.

For these conditions, i.e., positive integer values for $L$ fulfilling (\ref{NyquistCongChannelSamp}), it can be shown that the lower bound $\mathcal{R}_{L,\textrm{joint}}\big|_{\textrm{rect.} S_{h}(f)}$ in (\ref{PSE6}) is maximized for 
\begin{align}
L_{\textrm{opt}}&=\left\lfloor \frac{1}{2f_{d}}\right\rfloor . \label{L_optInt}
\end{align}

To prove this statement we differentiate the RHS of (\ref{PSE6}) with respect to $L$ and set the result equal to zero, which yields that the RHS of (\ref{PSE6}) has a unique local extremum at
\begin{align}
\tilde{L}_{\textrm{opt}}&= \frac{1}{2f_{d}}\frac{C_{\textrm{perf}}(\rho)\rho}{\rho-C_{\textrm{perf}}(\rho)}.\label{L_opt}
\end{align}
Numerical evaluation shows that the factor $\frac{C_{\textrm{perf}}(\rho)\rho}{\rho-C_{\textrm{perf}}(\rho)}$ is larger than one. As (\ref{L_opt}) is the only local extremum of the RHS of (\ref{PSE6}), and with the constraints on $L$ given by (\ref{NyquistCongChannelSamp}) and the fact that $L$ is an integer value, and considering that $\mathcal{R}_{L,\textrm{joint}}\big|_{\textrm{rect.} S_{h}(f)}$ monotonically increases with $L$ for $L<\tilde{L}_{\textrm{opt}}$ we can conclude that the lower bound is maximized by $L_{\textrm{opt}}$ in (\ref{L_optInt}). 

Substituting $L$ in (\ref{PSE6}) by $L_{\textrm{opt}}$ in (\ref{L_optInt}) yields a lower bound on the achievable rate with joint processing in case the pilot spacing can be arbitrarily chosen while fulfilling (\ref{NyquistCongChannelSamp}).

\section{Numerical Evaluation}\label{Sect_NumericalEvaluation}
\begin{figure}
\centering
 \psfrag{fd}[cc][tc][1.3]{$f_{d}$}
 \psfrag{bits}[cc][cc][1.2]{[bit/ channel use]}
 \psfrag{LBSep xxxxxxxxxxxxxxxxxx}[lc][lc][1.0]{UB separate process. (\ref{UpperBoundSD})}
 \psfrag{UBSep xxxxxxxxxxxxxxxxxx}[lc][lc][1.0]{LB separate process. (\ref{LowerBoundSD})}
 \psfrag{LBJoint xxxxxxxxxxxxxxxx}[lc][lc][1.0]{LB joint processing (\ref{PSE6})}
 \psfrag{LBJointLoptxxxxxxxxxxxxxxx}[lc][lc][1.0]{LB joint proc. $L_{\textrm{opt}}$(\ref{PSE6})/(\ref{L_optInt})}
 \psfrag{0dB}[cc][cc][1.15]{$0$\,dB}
 \psfrag{6dB}[cc][cc][1.15]{$6$\,dB}
 \psfrag{12dB}[cc][cc][1.15]{$12$\,dB}
    \includegraphics[width=0.8\columnwidth]{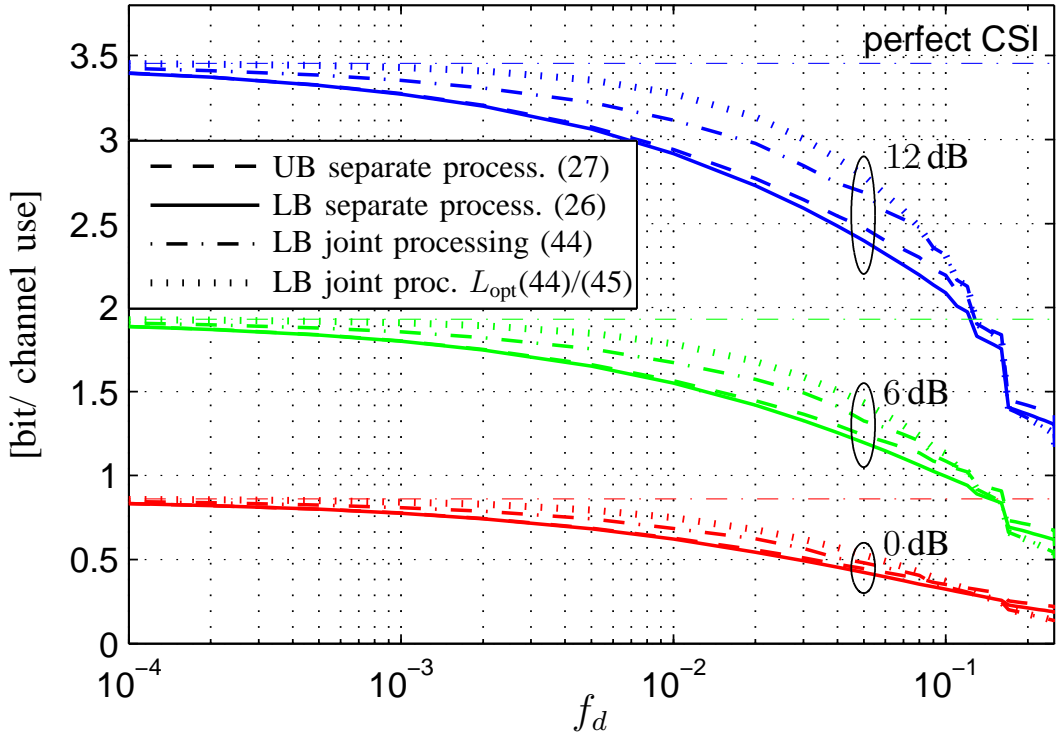}
    \caption{Comparison of bounds on the achievable rate with separate processing of data and pilot symbols to lower bounds on the achievable rate with joint processing of data and pilot symbols; except of \emph{LB joint proc. $L_{\textrm{opt}}$} the pilot spacing $L$ is chosen such that the lower bound for separate processing (\ref{LowerBoundSD}) is maximized; the PSD $S_{h}(f)$ is assumed to be rectangular, see (\ref{PSDRect})}
    \label{CompSepJoint_fixedL}
\end{figure}

Fig.~\ref{CompSepJoint_fixedL} shows a comparison of the bounds on the achievable rate for separate and joint processing of data and pilot symbols. 

On the one hand, the lower bound on the achievable rate for joint processing in (\ref{PSE6}) is compared to bounds on the achievable rate with separate processing of data and pilot symbols, i.e, (\ref{LowerBoundSD}) and (\ref{UpperBoundSD}), for a fixed pilot spacing. As the upper and lower bound on the achievable rate with separate processing are relatively tight, we choose the pilot spacing such that the lower bound on the achievable rate for separate processing in (\ref{LowerBoundSD}) is maximized. It can be seen that except for very high channel dynamics, i.e., very large $f_{d}$ the lower bound on the achievable rate for joint processing is larger than the bounds on the achievable rate with separate processing. This indicates the possible gain while using joint processing of data and pilot symbols for a given pilot spacing. Note, the observation that the lower bound for joint processing for large $f_{d}$ is smaller than the bounds on the achievable rate with separate processing is a result of the lower bounding, i.e., it indicates that the lower bound is not tight for these parameters.

On the other hand, also the lower bound on the achievable rate with joint processing and a pilot spacing that maximizes this lower bound, i.e., (\ref{PSE6}) in combination with (\ref{L_optInt}), is shown. In this case the pilot spacing is always chosen such that the channel fading process is sampled by the pilot symbols with Nyquist rate. Obviously, this lower bound is larger than or equal to the lower bound for joint processing while choosing the pilot spacing as it is optimal for separate processing of data and pilot symbols. This behavior arises from the effect that for separate processing in case of small $f_{d}$ a pilot rate is chosen that is higher than the Nyquist rate of the channel fading process to enhance the channel estimation quality. In case of a joint processing all symbols are used for channel estimation anyway. Therefore, a pilot rate higher than Nyquist rate always leads to an increased loss in the achievable rate as less symbols can be used for data transmission.

\begin{figure}
\centering
 \psfrag{fd}[cc][tc][1.3]{$f_{d}$}
 \psfrag{bits}[cc][cc][1.2]{[bit/ channel use]}
 \psfrag{LBIID xxxxxxxxxxxxxxxxx}[lc][lc][1.05]{UB i.i.d.\ PG (\ref{IIDUpp})}
 \psfrag{UBIID xxxxxxxxxxxxxxxxx}[lc][lc][1.05]{LB i.i.d.\ PG (\ref{IIDLow})}
 \psfrag{LBJointLoptxxxxxxxxxxxx}[lc][lc][1.05]{LB joint proc. $L_{\textrm{opt}}$ (\ref{PSE6})/(\ref{L_optInt})}
 \psfrag{0dB}[cc][cc][1.15]{$0$\,dB}
 \psfrag{6dB}[cc][cc][1.15]{$6$\,dB}
 \psfrag{12dB}[cc][cc][1.15]{$12$\,dB}
    \includegraphics[width=0.8\columnwidth]{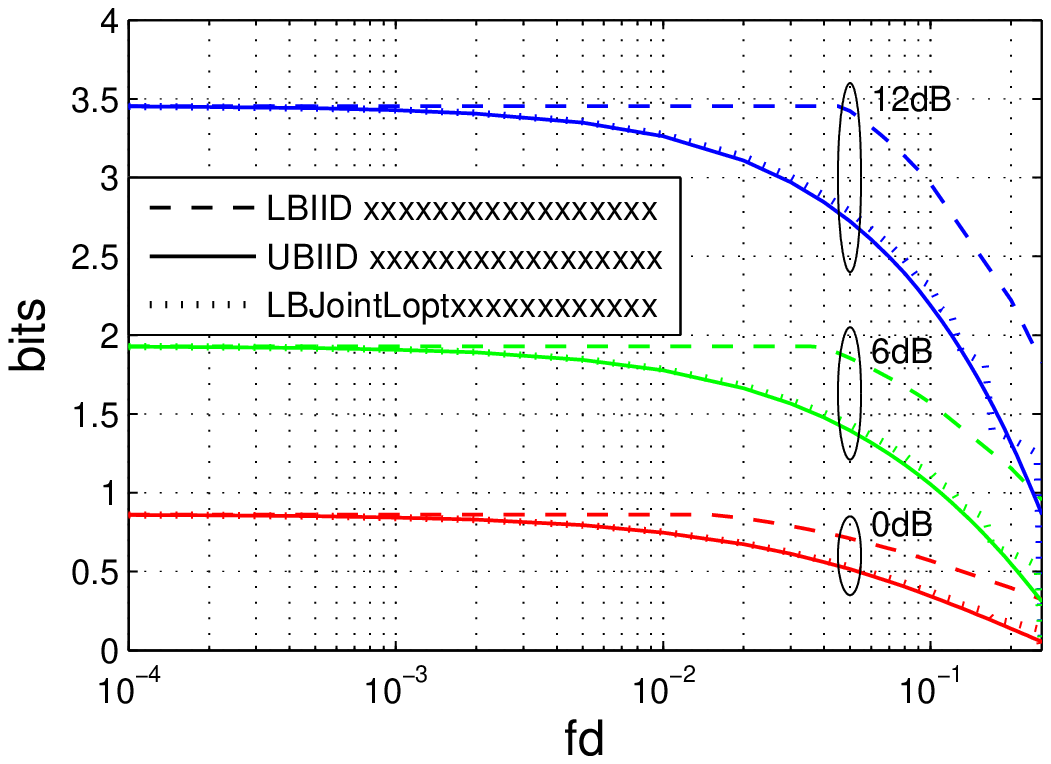}
    \caption{Lower bound on the achievable rate with joint processing of data and pilot symbols and a pilot spacing $L_{\textrm{opt}}$ that maximizes this bound, i.e., (\ref{PSE6}) in combination with (\ref{L_optInt}); for comparison bounds on the achievable rate with i.i.d.\ zero-mean proper Gaussian (PG) input symbols are shown; rectangular PSD $S_{h}(f)$, see (\ref{PSDRect})}
    \label{CompIIDJoint_arbitraryL}
\end{figure}

Fig.~\ref{CompIIDJoint_arbitraryL} shows the lower bound on the achievable rate for joint processing of data and pilot symbols when choosing $L$ as given in (\ref{L_optInt}), which maximizes the lower bound in (\ref{PSE6}). This lower bound is compared to the following bounds on the achievable rate with i.i.d.\ zero-mean proper Gaussian (PG) input symbols for a rectangular PSD of the channel fading process, see (\ref{PSDRect}), which have been given in \cite{Doer08ISITA}
\begin{align}
\mathcal{I}'_{L}(\mathbf{y};\mathbf{x})\big|_{\textrm{PG}}&=\max\left\{C_{\textrm{perf}}(\rho)-2f_{d}\log\left(1+\frac{\rho}{2f_{d}}\right),0\right\}\label{IIDLow}\\
\mathcal{I}'_{U}(\mathbf{y};\mathbf{x})\big|_{\textrm{PG}}&=\min\Bigg\{\log\left(1+\rho\right)-2f_{d}\int_{z=0}^{\infty}\log\left(1+\frac{\rho}{2f_{d}}z\right)e^{-z}dz, C_{\textrm{perf}}(\rho)\Bigg\}.\label{IIDUpp}
\end{align}
with $C_{\textrm{perf}}(\rho)$ being the coherent capacity of a Rayleigh flat-fading channel given in (\ref{CoherentCapacity}).

Obviously, for some parameters the lower bound on the achievable rate for joint processing of data and pilot symbols is larger than the lower bound on the achievable rate with i.i.d.\ zero-mean proper Gaussian input symbols, i.e., without the assumption of any pilot symbols. However, this observation does not allow to argue that in these cases the use of pilot symbols is better than i.i.d.\ symbols, as we only compare lower bounds.

\section{Summary}\label{Sect_Summary}
In the present work, we have studied the achievable rate with a joint processing of pilot and data symbols in the context of stationary Rayleigh flat-fading channels. We have discussed the nature of the possible gain when using joint processing of data and pilot symbols in contrast to separate processing. We have shown that the additional information that can be retrieved by joint processing is contained in the temporal correlation of the channel estimation error process when using a solely pilot based channel estimation, which cannot be captured by standard decoders as they are used in conventional synchronized detection based receivers with a solely pilot based channel estimation. In addition, and this is the main novelty of the present work, we have derived a lower bound on the achievable rate for joint processing of data and pilot symbols on a stationary Rayleigh flat-fading channel, giving an indication on the possible gain in terms of the achievable rate when using a joint processing of pilot and data symbols in comparison to the typically used separate processing.

\appendices
\section{Minimization of $h'(\mathbf{e}_{\mathrm{joint}}|\mathbf{x}_{D},\mathbf{x}_{P})$ by CM Modulation}\label{ProofLowerrBoundCM}
In this appendix we will show that the differential entropy rate $h'(\mathbf{e}_{\textrm{joint}}|\mathbf{x}_{D},\mathbf{x}_{P})$ in (\ref{DiffEntropyJointErrorRate_Joint}), which depends on the distribution of the data symbols contained in $\mathbf{x}_{D}$, is minimized for constant modulus input symbols among all distributions of the data symbols with an maximum average power of $\sigma_{x}^{2}$. 

The MAP channel estimate based on pilot and perfectly known data symbols is given by
\begin{align}
\hat{\mathbf{h}}_{\textrm{joint}}&=\arg\max_{\mathbf{h}}p(\mathbf{h}|\mathbf{y},\mathbf{x})\nonumber\\
&=\arg\max_{\mathbf{h}}p(\mathbf{y}|\mathbf{h},\mathbf{x})p(\mathbf{h})\nonumber\\
&=\arg\max_{\mathbf{h}}\left\{\log (p(\mathbf{y}|\mathbf{h},\mathbf{x}))+\log(p(\mathbf{h}))\right\}\label{MAPEstGen1}
\end{align}
with
\begin{align}
p(\mathbf{y}|\mathbf{h},\mathbf{x})&=\frac{1}{\pi^{N}\sigma_{n}^{2N}}\exp\left(-\frac{|\mathbf{y}-\mathbf{X}\mathbf{h}|^{2}}{\sigma_{n}^{2}}\right)\\
p(\mathbf{h})&=\frac{1}{\pi^{N}\det(\mathbf{R}_{h})}\exp\left(-\mathbf{h}^{H}\mathbf{R}_{h}^{-1}\mathbf{h}\right).
\end{align}
Thus, (\ref{MAPEstGen1}) becomes
\begin{align}
\hat{\mathbf{h}}_{\textrm{joint}}&=\arg\max_{\mathbf{h}}\left\{-\frac{1}{\sigma_{n}^{2}}|\mathbf{y}-\mathbf{X}\mathbf{h}|^{2}-\mathbf{h}^{H}\mathbf{R}_{h}^{-1}\mathbf{h}\right\}.\label{MAPEstGen2}
\end{align}
Differentiating the argument of the maximum operation at the RHS of (\ref{MAPEstGen2}) with respect to $\mathbf{h}$ and setting the result equal to zero yields
\begin{align}
&-\frac{1}{\sigma_{n}^{2}}\left\{-\mathbf{X}^{H}\mathbf{y}+\mathbf{X}^{H}\mathbf{X}\mathbf{h}\right\}-\mathbf{R}_{h}^{-1}\mathbf{h}=\mathbf{0}
\end{align}
and, thus,\footnote{Note that the inverse of $\mathbf{X}$ in (\ref{JointMAPEstimate}) does not exist, if a diagonal element of the diagonal matrix $\mathbf{X}$ is zero, i.e., one transmit symbol has zero power. However, as the channel estimates can be rewritten as
\begin{align}
\hat{\mathbf{h}}_{\textrm{joint}}&=\mathbf{R}_{h}\mathbf{X}^{H}\left(\mathbf{X}\mathbf{R}_{h}\mathbf{X}^{H}+\sigma_{n}^{2}\mathbf{I}_{N}\right)^{-1}\mathbf{y}
\end{align}
it is obvious that the elements of $\hat{\mathbf{h}}_{\textrm{joint}}$ are continuous in $x_{k}$ for all $k$, and, thus,  this does not lead to problems in the following derivation.} 
\begin{align}
\hat{\mathbf{h}}_{\textrm{joint}}&=\mathbf{R}_{h}\left(\mathbf{R}_{h}+\sigma_{n}^{2}\mathbf{X}^{-1}\mathbf{X}^{H^{-1}}\right)^{-1}\mathbf{X}^{-1}\mathbf{y}.\label{JointMAPEstimate}
\end{align}
With (\ref{JointMAPEstimate}) the channel estimation error correlation matrix $\mathbf{R}_{e_{\textrm{joint}}}$ is given by
\begin{align}
\mathbf{R}_{e_{\textrm{joint}}}&=\mathrm{E}\left[\left(\mathbf{h}-\hat{\mathbf{h}}_{\textrm{joint}}\right)\left(\mathbf{h}-\hat{\mathbf{h}}_{\textrm{joint}}\right)^{H}\Big|\mathbf{x}\right]\nonumber\\
&=\mathbf{R}_{h}-\mathbf{R}_{h}\left(\mathbf{R}_{h}+\sigma_{n}^{2}(\mathbf{X}^{H}\mathbf{X})^{-1}\right)^{-1}\mathbf{R}_{h}.\label{ChannelEstJointErrorCorr_Explicite}
\end{align}
Thus, the differential entropy $h(\mathbf{e}_{\textrm{joint}}|\mathbf{x}_{D},\mathbf{x}_{P})$ becomes
\begin{align}
h(\mathbf{e}_{\textrm{joint}}|\mathbf{x}_{D},\mathbf{x}_{P})&=\mathrm{E}_{\mathbf{x}}\left[\log\det\left(\pi e \mathbf{R}_{e_{\textrm{joint}}}\right)\right]\nonumber\\
&=\log\left((\pi e )^{N}\det(\mathbf{R}_{h})\right)+\mathrm{E}_{\mathbf{x}}\left[\log\det\left(\mathbf{I}_{N}-\left(\mathbf{R}_{h}+\sigma_{n}^{2}(\mathbf{X}^{H}\mathbf{X})^{-1}\right)^{-1}\mathbf{R}_{h}\right)\right].\label{AppendixDeriv1}
\end{align}
The argument of the expectation operation in the last summand on the RHS of (\ref{AppendixDeriv1}) can be rewritten as 
\begin{align}
&\log\det\left(\mathbf{I}_{N}-\left(\mathbf{R}_{h}+\sigma_{n}^{2}(\mathbf{X}^{H}\mathbf{X})^{-1}\right)^{-1}\mathbf{R}_{h}\right)\nonumber\\
&\qquad=\log\det\left(\mathbf{I}_{N}-\left(\mathbf{I}_{N}+\mathbf{R}_{h}^{-1}\sigma_{n}^{2}(\mathbf{X}^{H}\mathbf{X})^{-1}\right)^{-1}\right)\nonumber\\
&\qquad\stackrel{(a)}{=}\log\det\left(\mathbf{I}_{N}-\left[\mathbf{I}_{N}-\left(\frac{1}{\sigma_{n}^{2}}\mathbf{R}_{h}+(\mathbf{X}^{H}\mathbf{X})^{-1}\right)^{-1}(\mathbf{X}^{H}\mathbf{X})^{-1}\right]\right)\nonumber\\
&\qquad=-\log\det\left(\frac{1}{\sigma_{n}^{2}}\mathbf{R}_{h}\mathbf{X}^{H}\mathbf{X}+\mathbf{I}_{N}\right)\label{AppendixDeriv2}
\end{align}
where (a) follows from the matrix inversion lemma. Inserting (\ref{AppendixDeriv2}) into (\ref{AppendixDeriv1}) yields
\begin{align}
h(\mathbf{e}_{\textrm{joint}}|\mathbf{x}_{D},\mathbf{x}_{P})&=\log\left((\pi e )^{N}\det(\mathbf{R}_{h})\right)-\mathrm{E}_{\mathbf{x}}\left[\log\det\left(\frac{1}{\sigma_{n}^{2}}\mathbf{R}_{h}\mathbf{X}^{H}\mathbf{X}+\mathbf{I}_{N}\right)\right].\label{AppendixDeriv3}
\end{align}

As the matrix $\mathbf{X}=\textrm{diag}(\mathbf{x})$ is diagonal, the product $\mathbf{X}\mathbf{X}^{H}$ is also diagonal and its diagonal elements are the powers of the individual transmit symbols. In the following we substitute this product by
\begin{align}
\mathbf{Z}&=\mathbf{X}\mathbf{X}^{H}\label{DefJointAppZ_XX}
\end{align}
and $\mathbf{z}=\textrm{diag}(\mathbf{Z})$ contains the diagonal elements of $\mathbf{Z}$. 

The aim of this appendix is to show that the entropy rate $h'(\mathbf{e}_{\textrm{joint}}|\mathbf{x}_{D},\mathbf{x}_{P})$ corresponding to the entropy in (\ref{AppendixDeriv3}) is minimized by constant modulus data symbols with the power $\sigma_{x}^{2}$ among all input distributions fulfilling the maximum average power constraint in (\ref{AveragePowerConstraintJoint}), i.e., 
\begin{align}
\mathrm{E}\left[\mathbf{x}^{H}\mathbf{x}\right]&=\mathrm{E}\left[\sum_{k=1}^{N}z_{k}\right]\le N\sigma_{x}^{2}\label{AppendixDeriv4}
\end{align}
where the $z_{k}$ with $k=1\hdots N$ are the elements of $\mathbf{z}$. Therefor, in a first step, we study the entropy in (\ref{AppendixDeriv3}), i.e., a finite transmission length $N$. Let the set $\mathcal{P}$ be the set containing all input distributions fulfilling the maximum average power constraint in (\ref{AppendixDeriv4}). Note that this set $\mathcal{P}$ includes the case of having pilot symbols. However, when using pilot symbols, the transmit power of each $L$-th symbol is fixed to $\sigma_{x}^{2}$. For the moment, we allow all input distributions contained in $\mathcal{P}$. Later on, we will come back to the special case of using pilot symbols.

We want to find the input vector $\mathbf{z}$ that minimizes (\ref{AppendixDeriv3}) provided that the average power constraint is fulfilled. Therefor, we first show that the argument of the expectation operation on the RHS of (\ref{AppendixDeriv3}), i.e.,
\begin{align}
g(\mathbf{Z})=\log\det\left(\frac{1}{\sigma_{n}^{2}}\mathbf{R}_{h}\mathbf{Z}+\mathbf{I}_{N}\right)\label{DefgZ}
\end{align}
is concave in $\mathbf{Z}$. To verify the concavity of $g(\mathbf{Z})$, we follow along the lines of \cite[Chapter~3.1.5]{BoydConvOpt04} and consider an arbitrary line $\mathbf{Z}=\bar{\mathbf{Z}}+t\mathbf{\Delta}$. Based on this, we define $g(t)$ as
\begin{align}
g(t)&=\log\det\left(\frac{1}{\sigma_{n}^{2}}\mathbf{R}_{h}\left(\bar{\mathbf{Z}}+t\mathbf{\Delta}\right)+\mathbf{I}_{N}\right)\nonumber\\
&=\log\det\left(\frac{1}{\sigma_{n}^{2}}\mathbf{R}_{h}\right)+\log\det\left(\bar{\mathbf{Z}}+\sigma_{n}^{2}\mathbf{R}_{h}^{-1}+t  \mathbf{\Delta}\right)\nonumber\\
&\stackrel{(a)}{=}\log\det\left(\frac{1}{\sigma_{n}^{2}}\mathbf{R}_{h}\right)+\log\det\left(\mathbf{Q}+t \mathbf{\Delta}\right)\nonumber\\
&=\log\det\left(\frac{\mathbf{R}_{h}}{\sigma_{n}^{2}}\right)+\log\det\left(\mathbf{Q}^{\frac{H}{2}}\left(\mathbf{I}_{N}+t \mathbf{Q}^{-\frac{H}{2}}\mathbf{\Delta}\mathbf{Q}^{-\frac{1}{2}}\right)\mathbf{Q}^{\frac{1}{2}}\right)\nonumber\\
&=\log\det\left(\frac{\mathbf{R}_{h}}{\sigma_{n}^{2}}\right)+\log\det\left(\mathbf{Q}\right)+\log\det\left(\mathbf{I}_{N}+t \mathbf{Q}^{-\frac{H}{2}}\mathbf{\Delta}\mathbf{Q}^{-\frac{1}{2}}\right)\nonumber\\
&=\log\det\left(\frac{1}{\sigma_{n}^{2}}\mathbf{R}_{h}\bar{\mathbf{Z}}+\mathbf{I}_{N}\right)+\log\det\left(\mathbf{I}_{N}+t \left(\bar{\mathbf{Z}}+\sigma_{n}^{2}\mathbf{R}_{h}^{-1}\right)^{-\frac{H}{2}}\mathbf{\Delta}\left(\bar{\mathbf{Z}}+\sigma_{n}^{2}\mathbf{R}_{h}^{-1}\right)^{-\frac{1}{2}}\right)\nonumber\\
&\stackrel{(b)}{=}\log\det\left(\frac{1}{\sigma_{n}^{2}}\mathbf{R}_{h}\bar{\mathbf{Z}}+\mathbf{I}_{N}\right)+\sum_{k=1}^{N}\log\left(1+t\lambda_{k}\right)\label{Appgt}
\end{align}
where for (a) we have used the substitution $\mathbf{Q}\stackrel{\triangle}{=}\bar{\mathbf{Z}}+\sigma_{n}^{2}\mathbf{R}_{h}^{-1}$ to simplify notation. Furthermore, the $\lambda_{k}$ in (b) are the eigenvalues of $\left(\bar{\mathbf{Z}}+\sigma_{n}^{2}\mathbf{R}_{h}^{-1}\right)^{-\frac{H}{2}}\mathbf{\Delta}\left(\bar{\mathbf{Z}}+\sigma_{n}^{2}\mathbf{R}_{h}^{-1}\right)^{-\frac{1}{2}}$.

Based on (\ref{Appgt}) the derivatives of $g(t)$ with respect to $t$ are given by
\begin{align}
\frac{d g(t)}{dt}&=\sum_{k=1}^{N}\frac{\lambda_{k}}{1+t \lambda_{k}}\\
\frac{d^{2} g(t)}{dt^{2}}&=-\sum_{k=1}^{N}\frac{\lambda_{k}^{2}}{\left(1+t \lambda_{k}\right)^{2}}.
\end{align}
As the second derivative $\frac{d^{2} g(t)}{dt^{2}}$ is always negative, $g(\mathbf{Z})$ is concave on the set of diagonal matrices $\mathbf{Z}$ with non-negative diagonal entries. 

Based on the concavity of $g(\mathbf{Z})$ with respect to $\mathbf{Z}$ we can lower-bound $h(\mathbf{e}_{joint}|\mathbf{x}_{D},\mathbf{x}_{P})$ in (\ref{AppendixDeriv3}) by using Jensen's inequality as follows, cf. (\ref{DefgZ}):
\begin{align}
h(\mathbf{e}_{\textrm{joint}}|\mathbf{x}_{D},\mathbf{x}_{P})&=\log\det\left((\pi e)^{N}\det(\mathbf{R}_{h})\right)-\mathrm{E}_{\mathbf{z}}\left[g(\mathbf{Z})\right]\nonumber\\
&\ge \log\det\left((\pi e)^{N}\det(\mathbf{R}_{h})\right)-\log\det\left(\frac{1}{\sigma_{n}^{2}}\mathbf{R}_{h}\mathrm{E}\left[\mathbf{Z}\right]+\mathbf{I}_{N}\right).\label{AppendixErrorEntropMinJoint}
\end{align}

Recall, that we want to show that constant modulus data symbols with the power $\sigma_{x}^{2}$ minimize the entropy rate $h'(\mathbf{e}_{\textrm{joint}}|\mathbf{x}_{D},\mathbf{x}_{P})$. Therefore, from here on we consider the entropy rate which is given by
\begin{align}
h'(\mathbf{e}_{\textrm{joint}}|\mathbf{x}_{D},\mathbf{x}_{P})&=\lim_{N\rightarrow \infty}\frac{1}{N}h(\mathbf{e}_{\textrm{joint}}|\mathbf{x}_{D},\mathbf{x}_{P})\nonumber\\
&=\lim_{N\rightarrow \infty}\frac{1}{N}\left[\log\det\left((\pi e)^{N}\det(\mathbf{R}_{h})\right)-\log\det\left(\frac{1}{\sigma_{n}^{2}}\mathbf{R}_{h}\mathrm{E}\left[\mathbf{Z}\right]+\mathbf{I}_{N}\right)\right].\label{AppendixErrorEntropMinJointRate}
\end{align}

In the next step, we show for which kind of distribution of $\mathbf{z}$ fulfilling the maximum average power constraint in (\ref{AppendixDeriv4}) the RHS of (\ref{AppendixErrorEntropMinJointRate}) is minimized. I.e., we have to find
\begin{align}
&\lim_{N\rightarrow\infty}\frac{1}{N}\sup_{\mathcal{P}}\log\det\left(\frac{1}{\sigma_{n}^{2}}\mathbf{R}_{h}\mathrm{E}\left[\mathbf{Z}\right]+\mathbf{I}_{N}\right)\label{DefAppJointSup}
\end{align}
where the set $\mathcal{P}$ contains all input distributions fulfilling the maximum average power constraint in (\ref{AppendixDeriv4}). 

For the evaluation of (\ref{DefAppJointSup}) we substitute the Toeplitz matrix $\mathbf{R}_{h}$ by an asymptotic equivalent circulant matrix $\mathbf{C}_{h}$, which is possible, as we are finally interested in the supremum in (\ref{DefAppJointSup}) for the case of an infinite transmission length, i.e., $N\rightarrow \infty$. In the following, we will formalize the construction of $\mathbf{C}_{h}$ and show that the following holds
\begin{align}
\lim_{N\rightarrow\infty}\frac{1}{N}\sup_{\mathcal{P}}\log\det\left(\frac{1}{\sigma_{n}^{2}}\mathbf{R}_{h}\mathrm{E}\left[\mathbf{Z}\right]+\mathbf{I}_{N}\right)
&=\lim_{N\rightarrow\infty}\frac{1}{N}\sup_{\mathcal{P}}\log\det\left(\frac{1}{\sigma_{n}^{2}}\mathbf{C}_{h}\mathrm{E}\left[\mathbf{Z}\right]+\mathbf{I}_{N}\right)\label{AppJointMinEntropErr_CircToepEquiv_Main}
\end{align}

Therefore, we express the channel correlation matrix $\mathbf{R}_{h}$ by its spectral decomposition 
\begin{align}
\mathbf{R}_{h}&=\mathbf{R}_{h}^{(N)}=\mathbf{U}^{(N)}\mathbf{\Lambda}_{h}^{(N)}\left(\mathbf{U}^{(N)}\right)^{H}
\end{align}
where we introduced the superscript $(N)$ to indicate the size of the matrices. Furthermore, the matrix $\mathbf{U}^{(N)}$ is unitary and $\mathbf{\Lambda}_{h}^{(N)}=\textrm{diag}(\lambda_{1}^{(N)},\hdots,\lambda_{N}^{(N)})$ is diagonal and contains the eigenvalues $\lambda_{k}^{(N)}$ of $\mathbf{R}_{h}^{(N)}$.

We construct the circulant matrix $\mathbf{C}_{h}^{(N)}$ which is asymptotically equivalent to the Toeplitz matrix $\mathbf{R}_{h}^{(N)}$ following along the lines of \cite[Section 4.4, Eq. (4.32)]{Gray_ToeplitzReview}. The first column of the circulant matrix $\mathbf{C}_{h}^{(N)}$ is given by $(c_{0}^{(N)}, c_{1}^{(N)}, \hdots, c_{N-1}^{(N)})^{T}$ with the elements
\begin{align}
c_{k}^{(N)}&=\frac{1}{N}\sum_{l=0}^{N-1}\tilde{S}_{h}\left(\frac{l}{N}\right)e^{j2\pi \frac{l k}{N}}.
\end{align}
Here $\tilde{S}_{h}(f)$ is the periodic continuation of $S_{h}(f)$ given in (\ref{DefPSDH}), i.e., 
\begin{align}
\tilde{S}_{h}(f)&=\sum_{k=-\infty}^{\infty}\delta(f-k)\star S_{h}(f)\label{defineContPSD}
\end{align}
and $S_{h}(f)$ being zero outside the interval $|f|\le 0.5$ for which it is defined in (\ref{DefPSDH}). The asterisk $\star$ in (\ref{defineContPSD}) denotes convolution. 

As we assume that the autocorrelation function of the channel fading process is absolutely summable, see (\ref{ReqAbsSum}), the PSD of the channel fading process $\tilde{S}_{h}(f)$ is Riemann integrable, and it holds that
\begin{align}
\lim_{N\rightarrow \infty}c_{k}^{(N)}&=\lim_{N\rightarrow \infty}\frac{1}{N}\sum_{l=0}^{N-1}\tilde{S}_{h}\left(\frac{l}{N}\right)e^{j2\pi \frac{l k}{N}}\nonumber\\
&=\int_{0}^{1}\tilde{S}_{h}(f)e^{j2\pi k f}df\nonumber\\
&=\int_{-\frac{1}{2}}^{\frac{1}{2}}S_{h}(f)e^{j2\pi k f}df=r_{h}(k)
\end{align}
with $r_{h}(k)$ defined in (\ref{DefAutoCorrPrim}). 

As the eigenvectors of a circulant matrix are given by a discrete Fourier transform (DFT), the eigenvalues $\breve{\lambda}_{k}^{(N)}$ with $k=1,\hdots,N$ of the circulant matrix $\mathbf{C}_{h}^{(N)}$ are given by
\begin{align}
\breve{\lambda}_{k}^{(N)}&=\sum_{l=0}^{N-1}c_{l}^{(N)}e^{-j2\pi \frac{(k-1) l}{N}}\nonumber\\
&=\sum_{l=0}^{N-1}\left( \frac{1}{N}\sum_{m=0}^{N-1}\tilde{S}_{h}\left(\frac{m}{N}\right)e^{j2\pi \frac{m l}{N}}\right)e^{-j2\pi \frac{l (k-1)}{N}}\nonumber\\
&=\sum_{m=0}^{N-1}\tilde{S}_{h}\left(\frac{m}{N}\right)\left\{\frac{1}{N}\sum_{l=0}^{N-1}e^{j2\pi \frac{l(m-(k-1))}{N}}\right\}\nonumber\\
&=\tilde{S}_{h}\left(\frac{k-1}{N}\right).\label{DefEigenvalCirculant_EquivAsym}
\end{align}

Consequently, the spectral decomposition of the circulant matrix $\mathbf{C}_{h}^{(N)}$ is given by 
\begin{align}
\mathbf{C}_{h}^{(N)}&=\mathbf{F}^{(N)}\breve{\mathbf{\Lambda}}_{h}^{(N)}\left(\mathbf{F}^{(N)}\right)^{H}\label{SpectralDecompCirc}
\end{align}
where the matrix $\mathbf{F}^{(N)}$ is a unitary DFT matrix, i.e., its elements are given by
\begin{align}
\left[\mathbf{F}^{(N)}\right]_{k,l}&=\frac{1}{\sqrt{N}}e^{j2\pi\frac{(k-1)(l-1)}{N}}.\label{DefDFT}
\end{align}
Furthermore, the matrix $\breve{\mathbf{\Lambda}}_{h}^{(N)}$ is diagonal with the elements $\breve{\lambda}_{k}^{(N)}$ given in (\ref{DefEigenvalCirculant_EquivAsym}). 

By this construction the circulant matrix $\mathbf{C}_{h}^{(N)}$ is asymptotically equivalent to the Toeplitz matrix $\mathbf{R}_{h}^{(N)}$, see \cite[Lemma 4.6]{Gray_ToeplitzReview}, if the autocorrelation function $r_{h}(k)$ is absolutely summable, which is assumed to be fulfilled, see (\ref{ReqAbsSum}).

In the context of proving \cite[Lemma 4.6]{Gray_ToeplitzReview}, it is shown that the weak norm of the difference of $\mathbf{R}_{h}^{(N)}$ and $\mathbf{C}_{h}^{(N)}$ converges to zero as $N\rightarrow \infty$, i.e., 
\begin{align}
\lim_{N\rightarrow \infty}\left|\mathbf{R}_{h}^{(N)}-\mathbf{C}_{h}^{(N)}\right|&=0\label{WeakAsympEquivCircToplitz}
\end{align}
where the weak norm of a matrix $\mathbf{B}$ is defined as
\begin{align}
|\mathbf{B}|&=\left(\frac{1}{N}\textrm{Tr}\left[\mathbf{B}^{H}\mathbf{B}\right]\right)^{\frac{1}{2}}.
\end{align}
This fact will be used later on.

To exploit the asymptotic equivalence of $\mathbf{R}_{h}^{(N)}$ and $\mathbf{C}_{h}^{(N)}$ for the current problem, we have to show that the matrices in the argument of the $\log\det$ operation on the LHS and the RHS of (\ref{AppJointMinEntropErr_CircToepEquiv_Main}), i.e.,
\begin{align}
\mathbf{K}_{1}^{(N)}&=\frac{1}{\sigma_{n}^{2}}\mathbf{R}_{h}^{(N)}\mathrm{E}\left[\mathbf{Z}\right]+\mathbf{I}_{N}\\
\mathbf{K}_{2}^{(N)}&=\frac{1}{\sigma_{n}^{2}}\mathbf{C}_{h}^{(N)}\mathrm{E}\left[\mathbf{Z}\right]+\mathbf{I}_{N}
\end{align}
are asymptotically equivalent. 

In this context, we have to show that both matrices are bounded in the strong norm, and the weak norm of their difference converges to zero for $N\rightarrow \infty$ \cite[Section 2.3]{Gray_ToeplitzReview}. 

Concerning the condition with respect to the strong norm we have to show that
\begin{align}
\left\Vert \mathbf{K}_{1}^{(N)}\right\Vert =\left\Vert\frac{1}{\sigma_{n}^{2}}\mathbf{R}_{h}^{(N)}\mathrm{E}\left[\mathbf{Z}\right]+\mathbf{I}_{N}\right\Vert &< \infty\\
\left\Vert \mathbf{K}_{2}^{(N)}\right\Vert =\left\Vert \frac{1}{\sigma_{n}^{2}}\mathbf{C}_{h}^{(N)}\mathrm{E}\left[\mathbf{Z}\right]+\mathbf{I}_{N}\right\Vert &< \infty
\end{align}
with the strong norm of the matrix $\mathbf{B}$ defined by
\begin{align}
\left\Vert\mathbf{B}\right\Vert^{2} &=\max_{k}\gamma_{k}
\end{align}
where $\gamma_{k}$ are the eigenvalues of the Hermitian nonnegative definite matrix $\mathbf{B}\mathbf{B}^{H}$.
The diagonal matrix $\mathrm{E}\left[\mathbf{Z}\right]$ contains the average transmit powers of the individual transmit symbols on its diagonal. Thus, its entries are bounded. In addition, as the strong norms of $\mathbf{R}_{h}^{(N)}$ and $\mathbf{C}_{h}^{(N)}$ are bounded, too, the strong norms of $\mathbf{K}_{1}^{(N)}$ and $\mathbf{K}_{2}^{(N)}$ are bounded. Concerning the boundedness of the eigenvalues of the Hermitian Toeplitz matrix $\mathbf{R}_{h}^{(N)}$ see \cite[Lemma 4.1]{Gray_ToeplitzReview}.

Furthermore, the weak norm of the difference $\mathbf{K}_{1}^{(N)}-\mathbf{K}_{2}^{(N)}$ converges to zero for $N\rightarrow \infty$ as
\begin{align}
\left|\mathbf{K}_{1}^{(N)}-\mathbf{K}_{2}^{(N)}\right|&=\left|\frac{1}{\sigma_{n}^{2}}\mathbf{R}_{h}^{(N)}\mathrm{E}\left[\mathbf{Z}\right]+\mathbf{I}_{N}-\frac{1}{\sigma_{n}^{2}}\mathbf{C}_{h}^{(N)}\mathrm{E}\left[\mathbf{Z}\right]-\mathbf{I}_{N}\right|\nonumber\\
&=\left|\frac{1}{\sigma_{n}^{2}}\left(\mathbf{R}_{h}^{(N)}-\mathbf{C}_{h}^{(N)}\right)\mathrm{E}\left[\mathbf{Z}\right]\right|\nonumber\\
&\stackrel{(a)}{\le} \frac{1}{\sigma_{n}^{2}}\left|\mathbf{R}_{h}^{(N)}-\mathbf{C}_{h}^{(N)}\right| \Vert\mathrm{E}\left[\mathbf{Z}\right]\Vert 
\end{align}
where for (a) we have used \cite[Lemma 2.3]{Gray_ToeplitzReview}. As $\Vert\mathrm{E}\left[\mathbf{Z}\right]\Vert$  is bounded, we get for $N\rightarrow \infty$ 
\begin{align}
\lim_{N\rightarrow \infty}\left|\mathbf{K}_{1}^{(N)}-\mathbf{K}_{2}^{(N)}\right|&\le \lim_{N\rightarrow \infty}\frac{1}{\sigma_{n}^{2}}\left|\mathbf{R}_{h}^{(N)}-\mathbf{C}_{h}^{(N)}\right| \Vert\mathrm{E}\left[\mathbf{Z}\right]\Vert
=0
\end{align}
due to (\ref{WeakAsympEquivCircToplitz}). Thus we have shown the asymptotic equivalence of $\mathbf{K}_{1}^{(N)}$ and $\mathbf{K}_{2}^{(N)}$.

As $\mathbf{K}_{1}^{(N)}$ and $\mathbf{K}_{2}^{(N)}$ are asymptotically equivalent, with \cite[Theorem~2.4]{Gray_ToeplitzReview} the equality in (\ref{AppJointMinEntropErr_CircToepEquiv_Main}) holds. For ease of notation, in the following we omit the use of the superscript $(N)$ for all matrices and eigenvalues. 

Based on (\ref{AppJointMinEntropErr_CircToepEquiv_Main}) the evaluation of the supremum in (\ref{DefAppJointSup}) can be substituted by
\begin{align}
\lim_{N\rightarrow\infty}\frac{1}{N}\sup_{\mathcal{P}}\log\det\left(\frac{1}{\sigma_{n}^{2}}\mathbf{C}_{h}\mathrm{E}\left[\mathbf{Z}\right]+\mathbf{I}_{N}\right)
&\stackrel{(a)}{=}\lim_{N\rightarrow\infty}\frac{1}{N}\sup_{\mathcal{P}}\log\det\left(\frac{1}{\sigma_{n}^{2}}\mathbf{F}\breve{\mathbf{\Lambda}}_{h}\mathbf{F}^{H}\mathrm{E}\left[\mathbf{Z}\right]+\mathbf{I}_{N}\right)\nonumber\\
&\stackrel{(b)}{=}\lim_{N\rightarrow\infty}\frac{1}{N}\sup_{\mathcal{P}}\log\det\left(\frac{1}{\sigma_{n}^{2}}\breve{\mathbf{\Lambda}}_{h}\mathbf{F}^{H}\mathrm{E}\left[\mathbf{Z}\right]\mathbf{F}+\mathbf{I}_{N}\right)\label{JointAppProofMimiEntropErr_deriv1}
\end{align}
where for (a) we have used (\ref{SpectralDecompCirc}) and (b) is based on the following relation
\begin{align}
\det\left(\mathbf{A}\mathbf{B}+\mathbf{I}\right)&=\det\left(\mathbf{B}\mathbf{A}+\mathbf{I}\right)
\end{align}
which holds as $\mathbf{A}\mathbf{B}$ has the same eigenvalues as $\mathbf{B}\mathbf{A}$ for $\mathbf{A}$ and $\mathbf{B}$ being square matrices \cite[Theorem~1.3.20]{Horn}.

As the matrix $\frac{1}{\sigma_{n}^{2}}\breve{\mathbf{\Lambda}}_{h}\mathbf{F}^{H}\mathrm{E}\left[\mathbf{Z}\right]\mathbf{F}+\mathbf{I}_{N}$ in the argument of the logarithm on the RHS of (\ref{JointAppProofMimiEntropErr_deriv1}) is positive definite, using Hadamard's inequality we can upper-bound the argument of the supremum on the RHS of (\ref{JointAppProofMimiEntropErr_deriv1}) as follows
\begin{align}
\log\det\left(\frac{1}{\sigma_{n}^{2}}\breve{\mathbf{\Lambda}}_{h}\mathbf{F}^{H}\mathrm{E}\left[\mathbf{Z}\right]\mathbf{F}+\mathbf{I}_{N}\right)&\le \sum_{k=1}^{N}\log\left(\frac{1}{\sigma_{n}^{2}}\breve{\lambda}_{k}\left[\mathbf{F}^{H}\mathrm{E}\left[\mathbf{Z}\right]\mathbf{F}\right]_{k,k}+1\right)\label{ProofCMMinEntropJointAppHadamard}
\end{align}
where $\left[\mathbf{F}^{H}\mathrm{E}\left[\mathbf{Z}\right]\mathbf{F}\right]_{k,k}$ are the diagonal entries of the matrix $\mathbf{F}^{H}\mathrm{E}\left[\mathbf{Z}\right]\mathbf{F}$. Note, this means that distributions of the input sequences $\mathbf{z}$ which lead to the case that the matrix $\mathbf{F}^{H}\mathrm{E}\left[\mathbf{Z}\right]\mathbf{F}$ is diagonal maximize the RHS of (\ref{JointAppProofMimiEntropErr_deriv1}). Using (\ref{ProofCMMinEntropJointAppHadamard}), the RHS of (\ref{JointAppProofMimiEntropErr_deriv1}) is given by
\begin{align}
&\lim_{N\rightarrow\infty}\frac{1}{N}\sup_{\mathcal{P}}\log\det\left(\frac{1}{\sigma_{n}^{2}}\breve{\mathbf{\Lambda}}_{h}\mathbf{F}^{H}\mathrm{E}\left[\mathbf{Z}\right]\mathbf{F}+\mathbf{I}_{N}\right)\nonumber\\
&\qquad=\lim_{N\rightarrow\infty}\frac{1}{N}\sup_{\mathcal{P}}\sum_{k=1}^{N}\log\left(\frac{1}{\sigma_{n}^{2}}\breve{\lambda}_{k}\left(\frac{1}{N}\sum_{l=1}^{N}\mathrm{E}\left[z_{l}\right]\right)+1\right)\nonumber\\
&\qquad=\lim_{N\rightarrow\infty}\frac{1}{N}\sup_{\mathcal{P}}\sum_{k=1}^{N}\log\left(\frac{1}{\sigma_{n}^{2}}\breve{\lambda}_{k}\left(\mathrm{E}\left[\frac{1}{N}\sum_{l=1}^{N}z_{l}\right]\right)+1\right).\label{DeivJointMinEntropJointCM2}
\end{align}

It rests to evaluate the supremum on the RHS of (\ref{DeivJointMinEntropJointCM2}). However, as the logarithm is a monotonically increasing function with the maximum average power constraint in (\ref{AppendixDeriv4}) the supremum in (\ref{DeivJointMinEntropJointCM2}) is given by
\begin{align}
\lim_{N\rightarrow\infty}\frac{1}{N}\sup_{\mathcal{P}}\sum_{k=1}^{N}\log\left(\frac{1}{\sigma_{n}^{2}}\breve{\lambda}_{k}\left(\mathrm{E}\left[\frac{1}{N}\sum_{l=1}^{N}z_{l}\right]\right)+1\right)&=\lim_{N\rightarrow\infty}\frac{1}{N}\sum_{k=1}^{N}\log\left(\frac{\sigma_{x}^{2}}{\sigma_{n}^{2}}\breve{\lambda}_{k}+1\right)\nonumber\\
&\stackrel{(a)}{=}\lim_{N\rightarrow\infty}\frac{1}{N}\log\det\left(\frac{\sigma_{x}^{2}}{\sigma_{n}^{2}}\mathbf{C}_{h}+1\right)\nonumber\\
&\stackrel{(b)}{=}\lim_{N\rightarrow\infty}\frac{1}{N}\log\det\left(\frac{\sigma_{x}^{2}}{\sigma_{n}^{2}}\mathbf{R}_{h}+1\right)\label{DeivJointMinEntropJointCM3}
\end{align}
where (a) is based on (\ref{SpectralDecompCirc}) and for (b) we have used the asymptotic equivalence of the circulant matrix $\mathbf{C}_{h}$ and the Toeplitz matrix $\mathbf{R}_{h}$. 

Now, using (\ref{AppJointMinEntropErr_CircToepEquiv_Main}), (\ref{JointAppProofMimiEntropErr_deriv1}), (\ref{DeivJointMinEntropJointCM2}), and (\ref{DeivJointMinEntropJointCM3}) the supremum in (\ref{DefAppJointSup}) is given by
\begin{align}
\lim_{N\rightarrow\infty}\frac{1}{N}\sup_{\mathcal{P}}\log\det\left(\frac{1}{\sigma_{n}^{2}}\mathbf{R}_{h}\mathrm{E}\left[\mathbf{Z}\right]+\mathbf{I}_{N}\right)&=\lim_{N\rightarrow\infty}\frac{1}{N}\log\det\left(\frac{\sigma_{x}^{2}}{\sigma_{n}^{2}}\mathbf{R}_{h}+\mathbf{I}_{N}\right).
\end{align}
However, this means that the entropy rate $h'(\mathbf{e}_{\textrm{joint}}|\mathbf{x}_{D},\mathbf{x}_{P})$ in (\ref{AppendixErrorEntropMinJointRate}) is lower-bounded by
\begin{align}
h'(\mathbf{e}_{\textrm{joint}}|\mathbf{x}_{D},\mathbf{x}_{P})&=\lim_{N\rightarrow \infty}\frac{1}{N}\left[\log\det\left((\pi e)^{N}\det(\mathbf{R}_{h})\right)-\log\det\left(\frac{1}{\sigma_{n}^{2}}\mathbf{R}_{h}\mathrm{E}\left[\mathbf{Z}\right]+\mathbf{I}_{N}\right)\right]\nonumber\\
&\ge \lim_{N\rightarrow \infty}\frac{1}{N}\left[\log\det\left((\pi e)^{N}\det(\mathbf{R}_{h})\right)-\log\det\left(\frac{\sigma_{x}^{2}}{\sigma_{n}^{2}}\mathbf{R}_{h}+\mathbf{I}_{N}\right)\right]\nonumber\\
&\stackrel{(a)}{=}\lim_{N\rightarrow\infty}\frac{1}{N}\log\det\left(\pi e \mathbf{R}_{e_{\textrm{joint,CM}}}\right)\label{AlternativeIIDConcavity}
\end{align}
where for (a) we have used (\ref{AppendixDeriv1}) and (\ref{AppendixDeriv2}), and where $\mathbf{R}_{e_{\textrm{joint,CM}}}$ is the estimation error correlation matrix in case all input symbols have a constant modulus with power $\sigma_{x}^{2}$, cf. (\ref{ChannelEstJointErrorCorr_Explicite})
\begin{align}
\mathbf{R}_{e_{\textrm{joint,CM}}}&=\mathbf{R}_{h}-\mathbf{R}_{h}\left(\mathbf{R}_{h}+\frac{\sigma_{n}^{2}}{\sigma_{x}^{2}}\mathbf{I}_{N}\right)^{-1}\mathbf{R}_{h}.
\end{align}
This mean, that the entropy rate $h'(\mathbf{e}_{\textrm{joint}}|\mathbf{x}_{D},\mathbf{x}_{P})$ is minimized for the given maximum average power constraint in (\ref{AveragePowerConstraintJoint}) when all input symbols are constant modulus input symbols with power $\sigma_{x}^{2}$. Note that this includes the case that each $L$-th symbol is a pilot symbol with power $\sigma_{x}^{2}$ and all other symbols are constant modulus data symbols with power $\sigma_{x}^{2}$. 

In conclusion, we have shown that the differential entropy rate $h'(\mathbf{e}_{\textrm{joint}}|\mathbf{x}_{D},\mathbf{x}_{P})$ is minimized for constant modulus data symbols with power $\sigma_{x}^{2}$, i.e.,
\begin{align}
h'(\mathbf{e}_{\textrm{joint}}|\mathbf{x}_{D},\mathbf{x}_{P})&\ge h'(\mathbf{e}_{\textrm{joint}}|\mathbf{x}_{D},\mathbf{x}_{P})\big|_{\textrm{CM}}\nonumber\\
&=\lim_{N\rightarrow \infty}\frac{1}{N}\log\det\left(\pi e \mathbf{R}_{e_{\textrm{joint,CM}}}\right).\label{AppendixProofFinal_remark}
\end{align}

\section{Estimation Error Spectra $S_{e_{\mathrm{pil}}}(f)$ and $S_{e_{\mathrm{joint,CM}}}(f)$}\label{SectionPilBasedErrorSpec}
First, we calculate the PSD $S_{e_{\textrm{pil}}}(f)$ of the channel estimation error in case of a solely pilot based channel estimation. The channel estimation error in the frequency domain is given by
\begin{align}
E_{N}(e^{j2\pi f})&=\sum_{k=1}^{N}e_{\textrm{pil},k}\cdot e^{-j2\pi f k}\label{DefFiniteSizeErrorTransfrom}
\end{align}
where $e_{\textrm{pil},k}$ are the elements of the vector $\mathbf{e}_{\textrm{pil}}$. In the following we are interested in the case $N\rightarrow \infty$. As in this case the sum in (\ref{DefFiniteSizeErrorTransfrom}) does not exist, in the following we discuss\linebreak $\lim_{N\rightarrow \infty}\frac{1}{N} E_{N}(e^{j2\pi f})$, which can be expressed as follows
\begin{align}
\lim_{N\rightarrow \infty}\frac{1}{N}E_{N}(e^{j2\pi f})&\stackrel{(a)}{=}\lim_{N\rightarrow \infty}\frac{1}{N}\sum_{l=1}^{L}E_{N,l}(e^{j2\pi Lf})e^{-j2\pi l f}\nonumber\\
&\stackrel{(b)}{=}\lim_{N\rightarrow \infty}\frac{1}{N}\sum_{l=1}^{L}\Bigg[H_{N,l}(e^{j2\pi L f})-W_{l}(e^{j2\pi Lf})\frac{Y_{N,P}(e^{j2\pi L f})}{\sigma_{x}}\Bigg]e^{-j 2\pi l f}\nonumber\\
&\stackrel{(c)}{=}\lim_{N\rightarrow \infty}\frac{1}{N}\Bigg[H_{N}(e^{j2\pi f})-\sum_{l=1}^{L}W(e^{j2\pi Lf})e^{j 2\pi lf}\frac{Y_{N,P}(e^{j2\pi L f})}{\sigma_{x}}e^{-j 2\pi lf}\Bigg]\nonumber\\
&=\lim_{N\rightarrow \infty}\frac{1}{N}\left[H_{N}(e^{j2\pi f})-L \cdot W(e^{j2\pi Lf})\frac{Y_{N,P}(e^{j2\pi L f})}{\sigma_{x}}\right]\nonumber\\
&\stackrel{(d)}{=}\lim_{N\rightarrow \infty}\frac{1}{N}\Bigg[H_{N}(e^{j2\pi f})-L \cdot W(e^{j2\pi Lf})\left[H_{N,P}(e^{j2\pi L f})+\frac{N_{N,P}(e^{j2\pi L f})}{\sigma_{x}}\right]\Bigg].\label{PSE1}
\end{align}
For (a) we have used that the estimation error in frequency domain is the sum of the interpolation errors at the individual symbols time instances between the pilot symbols, where the temporal shift yields the phase shift of $2\pi l f$. Here $E_{N,l}(e^{j2\pi Lf})$ is the frequency transform of the estimation error at the symbol position with the distance $l$ to the next pilot symbols, i.e.,
\begin{align}
E_{N,l}(e^{j2\pi Lf})&=\sum_{k=1}^{\frac{N}{L}}e_{\textrm{pil},(k-1)L+1+l}\cdot e^{-j2\pi f kL}, \quad \textrm{for }l=0,\hdots ,L-1
\end{align}
where without loss of generality we assume that $N$ is an integer multiple of $L$ and that the transmit sequence starts with a pilot symbol. Equality (b) results from expressing $E_{N,l}(e^{j2\pi Lf})$ by the difference between the actual channel realization and the estimated channel realization at the different interpolation positions in time domain transferred to frequency domain. Here, without loss of generality, we assume that the pilot symbols are given by $\sigma_{x}$. Furthermore, $W_{l}(e^{j2\pi Lf})$ is the transfer function of the interpolation filter for the symbols at distance $l$ from the previous pilot symbol. Furthermore, $Y_{N,P}(e^{j2\pi L f})$ is the channel output at the pilot symbols time instance transferred to frequency domain. For (c) we have used that the sum of the phase shifted channel realizations in frequency domain at sampling rate $1/L$ corresponds to the frequency domain representation of the fading process at symbol rate. In addition, we have used that for $N\rightarrow \infty$ the interpolation filter transfer function $W_{l}(e^{j2\pi Lf})$, which is an MMSE interpolation filter, can be expressed as
\begin{align}
W_{l}(e^{j2\pi Lf})&=W(e^{j2\pi Lf})e^{j 2\pi lf}
\end{align}
i.e., the interpolation filter transfer functions for the individual time shifts are equal except of a phase shift. Finally, for (d) we have expressed $Y_{N,P}(e^{j2\pi L f})$ as the sum of the frequency domain representations of the fading process and the additive noise process. 

Based on (\ref{PSE1}) the PSD $S_{e_{\textrm{pil}}}(f)$ is given by
\begin{align}
S_{e_{\textrm{pil}}}(f)&=\lim_{N\rightarrow \infty} \frac{1}{N}\mathrm{E}\left[|E_{N}(e^{j2\pi f})|^{2}\right]\nonumber\\
&=\lim_{N\rightarrow \infty}\frac{1}{N}\mathrm{E}\Bigg[|H_{N}(e^{j2\pi f})|^{2}-L\cdot H_{N}(e^{j2\pi f})W^{*}(e^{j2\pi Lf})H_{N,P}^{*}(e^{j2\pi Lf})\nonumber\\&\qquad\qquad\qquad-L\cdot H_{N}^{*}(e^{j2\pi f})W(e^{j2\pi Lf})H_{N,P}(e^{j2\pi Lf})\nonumber\\&\qquad\qquad\qquad+L^{2} |W(e^{j2\pi Lf})|^{2}\left[|H_{N,P}(e^{j2\pi Lf})|^{2}+\left|\frac{N_{N,P}(e^{j2\pi L f})}{\sigma_{x}}\right|^{2}\right]\Bigg]\nonumber\\
&=S_{h}(e^{j2\pi f})-\lim_{N\rightarrow \infty}\frac{1}{N}\mathrm{E}\Bigg[ L\cdot W^{*}(e^{j2\pi Lf})\sum_{l=1}^{L}H_{N,l}(e^{j2\pi Lf})e^{-j2\pi lf}H_{N,P}^{*}(e^{j2\pi Lf})\nonumber\\&\quad+ L\cdot W(e^{j2\pi Lf})\sum_{l=1}^{L}H^{*}_{N,l}(e^{j2\pi Lf})e^{j2\pi lf}H_{N,P}(e^{j2\pi Lf})\Bigg]\nonumber\\
&\quad +L^{2}|W(e^{j2\pi Lf})|^{2}\left[\frac{1}{L}S_{h}(e^{j2\pi Lf})+\frac{1}{L}\frac{\sigma_{n}^{2}}{\sigma_{x}^{2}}\right]\nonumber\\
&=S_{h}(e^{j2\pi f})-L\cdot W^{*}(e^{j2\pi Lf})\sum_{l=1}^{L}\frac{1}{L} \cdot S_{h}(e^{j2\pi Lf})-L\cdot W(e^{j2\pi Lf})\sum_{l=1}^{L}\frac{1}{L} \cdot S^{*}_{h}(e^{j2\pi Lf})\nonumber\\&\quad+L^{2}|W(e^{j2\pi Lf})|^{2}\left[\frac{1}{L}S_{h}(e^{j2\pi Lf})+\frac{1}{L}\frac{\sigma_{n}^{2}}{\sigma_{x}^{2}}\right]\nonumber\\
&\stackrel{(a)}{=}S_{h}(e^{j2\pi f})-2L\cdot W(e^{j2\pi Lf}) S_{h}(e^{j2\pi Lf})+L|W(e^{j2\pi Lf})|^{2}\left[S_{h}(e^{j2\pi Lf})+\frac{\sigma_{n}^{2}}{\sigma_{x}^{2}}\right]\label{PSE2}
\end{align}
where for (a) we have used that $S_{h}(f)$ is real and, thus, the MMSE filter $W(e^{j2\pi Lf})$ is also real, see below.

The MMSE filter transfer function $W(e^{j2\pi Lf})$ is given by
\begin{align}
W(e^{j2\pi Lf})&=\frac{S_{h}(e^{j2\pi Lf})}{S_{h}(e^{j2\pi Lf})+\frac{\sigma_{n}^{2}}{\sigma_{x}^{2}}}=\frac{\frac{1}{L}S_{h}(e^{j2\pi f})}{\frac{1}{L}S_{h}(e^{j2\pi f})+\frac{\sigma_{n}^{2}}{\sigma_{x}^{2}}}\label{MMSEInterpolFreq}
\end{align}
where we have used that 
\begin{align}
S_{h}(e^{j2\pi Lf})&=\frac{1}{L}S_{h}(e^{j2\pi f}).\label{PSE3}
\end{align}

Inserting (\ref{MMSEInterpolFreq}) into (\ref{PSE2}) yields
\begin{align}
S_{e_{\textrm{pil}}}(f)&=S_{h}(e^{j2\pi f})-2L \frac{S_{h}(e^{j2\pi f})}{S_{h}(e^{j2\pi f})+L\frac{\sigma_{n}^{2}}{\sigma_{x}^{2}}} S_{h}(e^{j2\pi Lf})+L\left|\frac{S_{h}(e^{j2\pi f})}{S_{h}(e^{j2\pi f})+L\frac{\sigma_{n}^{2}}{\sigma_{x}^{2}}}\right|^{2}\left[S_{h}(e^{j2\pi Lf})+\frac{\sigma_{n}^{2}}{\sigma_{x}^{2}}\right]\nonumber\\
&\stackrel{(a)}{=}S_{h}(e^{j2\pi f})-2\cdot \frac{S_{h}(e^{j2\pi f})}{S_{h}(e^{j2\pi f})+L\frac{\sigma_{n}^{2}}{\sigma_{x}^{2}}} S_{h}(e^{j2\pi f})+\frac{\left|S_{h}(e^{j2\pi f})\right|^{2}}{S_{h}(e^{j2\pi f})+L\frac{\sigma_{n}^{2}}{\sigma_{x}^{2}}}\nonumber\\\displaybreak[3]
&=\frac{S_{h}(e^{j2\pi f})L\frac{\sigma_{n}^{2}}{\sigma_{x}^{2}}}{S_{h}(e^{j2\pi f})+L\frac{\sigma_{n}^{2}}{\sigma_{x}^{2}}}\nonumber\\
&=\frac{S_{h}(e^{j2\pi f})}{\frac{\rho}{L}\frac{S_{h}(e^{j2\pi f})}{\sigma_{h}^{2}}+1}\nonumber\\
&\stackrel{(b)}{=}\frac{S_{h}(f)}{\frac{\rho}{L}\frac{S_{h}(f)}{\sigma_{h}^{2}}+1}\label{PSE4_App}
\end{align}
where (a) results from (\ref{PSE3}) and for (b) we simplified the notation and substituted $e^{j2\pi f}$ by $f$ to get a consistent notation with (\ref{DefPSDH}).

The PSD $S_{e_{\textrm{joint,CM}}}(f)$ is then obviously given by setting $L=1$ in (\ref{PSE4_App}), i.e., 
\begin{align}
S_{e_{\textrm{joint,CM}}}(f)&=\frac{S_{h}(f)}{\rho\frac{S_{h}(f)}{\sigma_{h}^{2}}+1}\label{PSE5_App}
\end{align}
as all data symbols are assumed to be known and of constant modulus with power $\sigma_{x}^{2}$, cf. (\ref{LowerBoundJoint4_gen}).

\bibliographystyle{IEEEtran}
\bibliography{Bib_mod_IEEE}

\end{document}